\documentclass[twocolumn]{aastex61}
\usepackage{multirow}
\usepackage{booktabs}
\usepackage{array}
\usepackage{amsmath}
\def\aj{AJ}% % Astronomical Journal 
\def\araa{ARA\&A}% % Annual Review of Astron and Astrophys 
\def\apj{ApJ}% % Astrophysical Journal 
\def\apjl{ApJL}% % Astrophysical Journal, Letters 
\def\apjs{ApJS}% % Astrophysical Journal, Supplement 
\def\ao{Appl.~Opt.}% % Applied Optics 
\def\aap{A\&A}% % Astronomy and Astrophysics 
\def\aapr{A\&A~Rev.}% % Astronomy and Astrophysics Reviews 
\def\mnras{MNRAS}% % Monthly Notices of the RAS 
\def\pasp{PASP}% % Publications of the ASP 
\DeclareMathAlphabet{\mathsc}{OT1}{cmr}{m}{sc}
\def\testbx{bx}%
\DeclareRobustCommand{\ion}[2]{%
\relax\ifmmode
\ifx\testbx\f@series
{\mathbf{#1\,\mathsc{#2}}}\else
{\mathrm{#1\,\mathsc{#2}}}\fi
\else\textup{#1\,{\mdseries\textsc{#2}}}%
\fi}
\newcommand{\Hi} {\ion{H}{i}}

\newcommand{\ha} {\mbox{H$\alpha$}}
\newcommand{\hb} {\mbox{H$\beta$}}

\newcommand{\Oi} {\ion{O}{i}}

\newcommand{\Nai} {\ion{Na}{i}}

\newcommand{\Caii} {\ion{Ca}{ii}}

\newcommand{\sn}{ASASSN-16at}
\newcommand{\host}{UGC~08041}
\newcommand{\EpEpoch}{JD~2457406.42}

\newcommand{\iraf}{\texttt{IRAF}}
\newcommand{\daophot}{\textsc{daophot}}

\newcommand{\ebv}{\mbox{$E(B-V)$}}

\newcommand{\maghundred}{\mbox{mag (100 d)$ ^{-1} $}}
\newcommand{\msun}{\mbox{M$_{\odot}$}}

\newcommand{\kms}{\mbox{$\rm{\,km\,s^{-1}}$}}

\newcommand{\ergs}{\mbox{$\rm{\,erg\,s^{-1}}$}}

\newcommand{\nickel}{\mbox{$^{56}$Ni}}
\newcommand{\cobalt}{\mbox{$^{56}$Co}}

\newcommand{\ld}{\mbox{$\lambda$}}
\newcommand{\ldld}{\mbox{$\lambda\lambda$}}
\newcommand{\swift}{{\it Swift}}
\newcommand{\chandra}{{\it Chandra}}

\newcommand{\plm}{$\pm$}
 
\begin{document}

\title{Strongly Bipolar Inner Ejecta of the normal Type IIP supernova \sn %\footnote{This paper includes data gathered with the 6.5 meter Magellan Telescopes located at Las Campanas Observatory, Chile.}
} 
\author{Subhash Bose} \affil{Kavli Institute for Astronomy and Astrophysics, Peking University, Yi He Yuan Road 5, Hai Dian District, Beijing 100871, China.}
\author{Subo Dong} \affil{Kavli Institute for Astronomy and Astrophysics, Peking University, Yi He Yuan Road 5, Hai Dian District, Beijing 100871, China.}
\author{N. Elias-Rosa} \affil{Institute of Space Sciences (ICE, CSIC), Campus UAB, Carrer de Can Magrans s/n, 08193 Barcelona, Spain} \affil{Institut d’Estudis Espacials de Catalunya (IEEC), c/Gran Capit\'a 2-4, Edif. Nexus 201, 08034 Barcelona, Spain}
\author{B. J. Shappee} \affil{Institute for Astronomy, University of Hawaii, 2680 Woodlawn Drive, Honolulu, HI 96822, USA}
\author{David Bersier} \affil{Astrophysics Research Institute, Liverpool Science Park, 146 Brownlow Hill, Liverpool L3 5RF, UK 0000-0001-7485-3020}
\author{Stefano Benetti} \affil{INAF-Osservatorio Astronomico di Padova, Vicolo dell'Osservatorio 5, I-35122 Padova, Italy}
\author{M. D. Stritzinger} \affil{Department of Physics and Astronomy, Aarhus University, Ny Munkegade 120, DK-8000 Aarhus C, Denmark }
\author{D. Grupe} \affil{Department of Earth and Space Sciences, Morehead State University, Morehead, KY, 40351, USA}
\author{C. S. Kochanek} \affil{Department of Astronomy, The Ohio State University, 140 W. 18th Avenue, Columbus, OH 43210, USA.} \affil{Center for Cosmology and AstroParticle Physics (CCAPP), The Ohio State University, 191 W. Woodruff Avenue, Columbus, OH 43210, USA.}
\author{J. L. Prieto} \affil{N\'ucleo de Astronom\'ia de la Facultad de Ingenier\'ia y Ciencias, Universidad Diego Portales, Av. Ej \'ercito 441, Santiago, Chile} \affil{Millennium Institute of Astrophysics, Santiago, Chile.}
\author{Ping Chen} \affil{Kavli Institute for Astronomy and Astrophysics, Peking University, Yi He Yuan Road 5, Hai Dian District, Beijing 100871, China.}
\author{H. Kuncarayakti} \affil{Finnish Centre for Astronomy with ESO (FINCA), FI-20014 University of Turku, Finland.} \affil{Tuorla Observatory, Department of Physics and Astronomy, FI-20014 University of Turku, Finland}
\author{Seppo Mattila} \affil{Tuorla Observatory, Department of Physics and Astronomy, FI-20014 University of Turku, Finland}
\author{Antonia Morales-Garoffolo} \affil{Department of Applied Physics, University of C\'adiz, Campus of Puerto Real, E-11510 C\'adiz, Spain}
\author{Nidia Morrell} \affil{Carnegie Observatories, Las Campanas Observatory, Casilla 601, La Serena, Chile}
\author{F. Onori} \affil{Istituto di Astrofisica e Planetologia Spaziali, via Fosso del Cavaliere 100,I-00133 Rome, Italy}
\author{Thomas M Reynolds} \affil{Tuorla Observatory, Department of Physics and Astronomy, FI-20014 University of Turku, Finland}
\author{A. Siviero} \affil{Dipartimento di Fisica e Astronomia, Universit`a di Padova, via Marzolo 8, I-35131 Padova, Italy}
\author{Auni Somero} \affil{Tuorla Observatory, Department of Physics and Astronomy, FI-20014 University of Turku, Finland}
\author{K. Z. Stanek} \affil{Department of Astronomy, The Ohio State University, 140 W. 18th Avenue, Columbus, OH 43210, USA.} \affil{Center for Cosmology and AstroParticle Physics (CCAPP), The Ohio State University, 191 W. Woodruff Avenue, Columbus, OH 43210, USA.}
\author{Giacomo Terreran} \affil{Center for Interdisciplinary Exploration and Research in Astrophysics (CIERA) and Department of Physics and Astronomy, Northwestern University, Evanston, IL 60208}
\author{Todd A. Thompson} \affil{Department of Astronomy, The Ohio State University, 140 W. 18th Avenue, Columbus, OH 43210, USA.} \affil{Center for Cosmology and AstroParticle Physics (CCAPP), The Ohio State University, 191 W. Woodruff Avenue, Columbus, OH 43210, USA.} \affil{Institute for Advanced Study, 1 Einstein Dr, Princeton, New Jersey 08540}
\author{L. Tomasella} \affil{INAF-Osservatorio Astronomico di Padova, Vicolo dell'Osservatorio 5, I-35122 Padova, Italy}
\author{C. Ashall} \affil{Department of Physics, Florida State University, Tallahassee, FL 32306, USA}
\author{Christa Gall} \affil{Dark Cosmology Centre, Niels Bohr Institute, University of Copenhagen, Juliane Maries Vej 30, 2100 Copenhagen, Denmark}
\author{M. Gromadzki} \affil{Warsaw University Astronomical Observatory, Al. Ujazdowskie 4, 00-478 Warszawa, Poland}
\author{T. W.-S. Holoien} \affil{Carnegie Observatories, 813 Santa Barbara Street, Pasadena, CA 91101, USA}

\correspondingauthor{Subo Dong, Subhash Bose}
\email{dongsubo@pku.edu.cn, email@subhashbose.com}

\begin{abstract}
We report distinctly double-peaked $H\alpha$ and $H\beta$ emission lines in the late-time, nebular-phase spectra ($\gtrsim200$~days) of the otherwise normal at early phases ($\lesssim 100$~days) Type IIP supernova \sn\,(SN~2016X). 
Such distinctly double-peaked nebular Balmer lines have never been observed for a Type II~SN.
{The nebular-phase Balmer emission is driven by the radioactive $^{56}$Co decay, so the observed line-profile bifurcation suggests a strong bipolarity in the \nickel\ distribution or in the line-forming region of the inner ejecta.}
The strongly bifurcated blue- and red-shifted peaks are separated by $\sim3\times10^3 $\kms\,and are roughly symmetrically positioned with respect to the host-galaxy rest frame, implying that the inner ejecta are composed of two almost detached blobs.
The red peak progressively weakens relative to the blue peak, and disappears in the $740$~days spectrum.
One possible reason for the line-ratio evolution is increasing differential extinction from  continuous formation of dust within the envelope, which is also supported by the near-infrared flux excess that develops after $\sim100$~days.
\end{abstract}


% sec:intro
%______________________________________________________________________________________________

\section{Introduction} \label{sec:intro}

Hydrogen-rich, core-collapse supernovae (CCSNe), also known as type II SNe, originate from massive stars ($ M\ge8\msun $) which have retained most of their hydrogen content at the time of explosion. 
From spectropolarimetry studies, CCSNe ejecta are often found to show a significant degree of asymmetry \citep[see, e.g., the review by][]{2008ARA&A..46..433W}. In the early photospheric phase of SNe II, the inner ejecta is  mostly obscured by the thick and extended envelope of ionized hydrogen, which become increasingly transparent as the ejecta expands and the hydrogen recombines. The late-time, ``nebular phase" observations are particularly important to unveil the structure of the inner regions once the ejecta becomes optically thin. During the late-time ($\gtrsim100-150$\,d) light-curve tail, the optical radiation is primarily powered by the decay of radioactive \cobalt\ (the decay product of \nickel\ synthesized in the explosion), and the \nickel\ distribution in the ejecta can be reflected in the nebular Balmer emission line profiles \citep{2007AIPC..937..357C}, which can be a powerful probe of the explosion asymmetry. Despite decades of studies of SNe II, the exact mechanism driving the shock within the SN during the explosion is still under debate \citep[see, e.g.,][and references therein]{2012ARNPS..62..407J, 2015ApJ...801...90P, 2015ApJ...811...97K}, and non-sphericity and jets are often suspected to be critical to these explosions \citep[e.g.,][]{1999ApJ...524L.107K,2012ARNPS..62..407J,2017arXiv170408298P,2018arXiv181009074S}. Studying the asymmetry of the ejecta, especially the inner region, may provide important clues in understanding the explosion mechanism.

CCSNe may be an important source of dust in the universe \citep[see, e.g.,][]{2011A&ARv..19...43G}. During the nebular phase, as the ejecta cools, the gas may start to condense into dust grains and thereby increase the extinction locally. Dust formation has been observed in several CCSNe \citep[e.g.,][]{2006Sci...313..196S,2008MNRAS.389..141M,2009ApJ...704..306K,2011ApJ...732..109M,2011MNRAS.417..261I,2012ApJ...756..173S,2013ApJ...776....5M}. The effect of newly formed dust can be manifested in the light curves and spectra of the SN. The dust absorbs the light and re-emits in infra-red (IR) resulting in strong IR excess. Substantial dust formation may also lead to asymmetries in nebular emission lines due to differential extinction as the light coming from the far side of the ejecta suffers more extinction.

Here we present detailed observations of the normal type IIP SN \sn\ until its late nebular phase (up to $\sim900\,$days). \sn\ shows a unique double-peaked profile in \ha\ and \hb\ nebular emission lines, where the relative strengths of the two peaks evolve with time. Such a distinct double-peaked structure is unprecedented for an SN IIP.  \cite{2018MNRAS.475.3959H} studied the NUV-optical light curves and optical spectra of \sn\ until the early radioactive decay tail phase, showing key features that are typical for a SN IIP. They also reported that, the \ha\,emission profile showed weak asymmetry in their last spectrum at 142d, for which they suggested three possible interpretations -- circumstellar medium (CSM) interaction, asymmetry in the line-emitting region or bipolar \nickel\ distribution. Our still later nebular-phase spectra taken at $\gtrsim200$\,d show Balmer lines with double-peaked profiles, which most likely suggest a bipolar distribution of the inner ejecta. Our work focuses on the nebular data, while our full photometry and spectroscopic data are given in the Appendix.

\begin{figure*}
	\centering
	\includegraphics[width=1\linewidth]{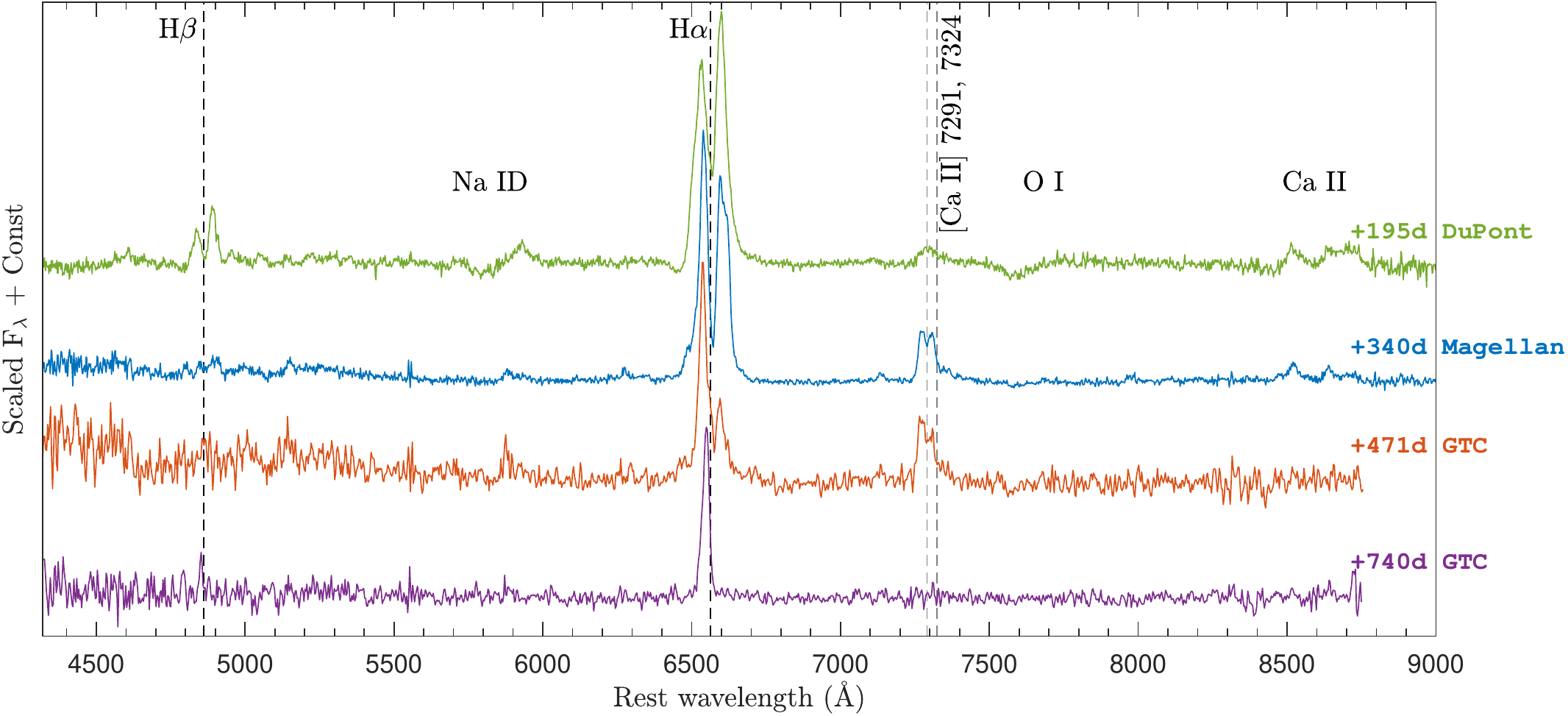}
	\caption{Deep nebular spectra of \sn\ at 4 epochs (195\,d, 340\,d, 471\,d and 740\,d). The vertical dashed lines mark the rest wavelengths of the \ha, \hb\ and the [\Caii] doublet.}
	\label{fig:spec_nebular}
\end{figure*}

\begin{figure}
	\centering
	\includegraphics[width=0.68\linewidth]{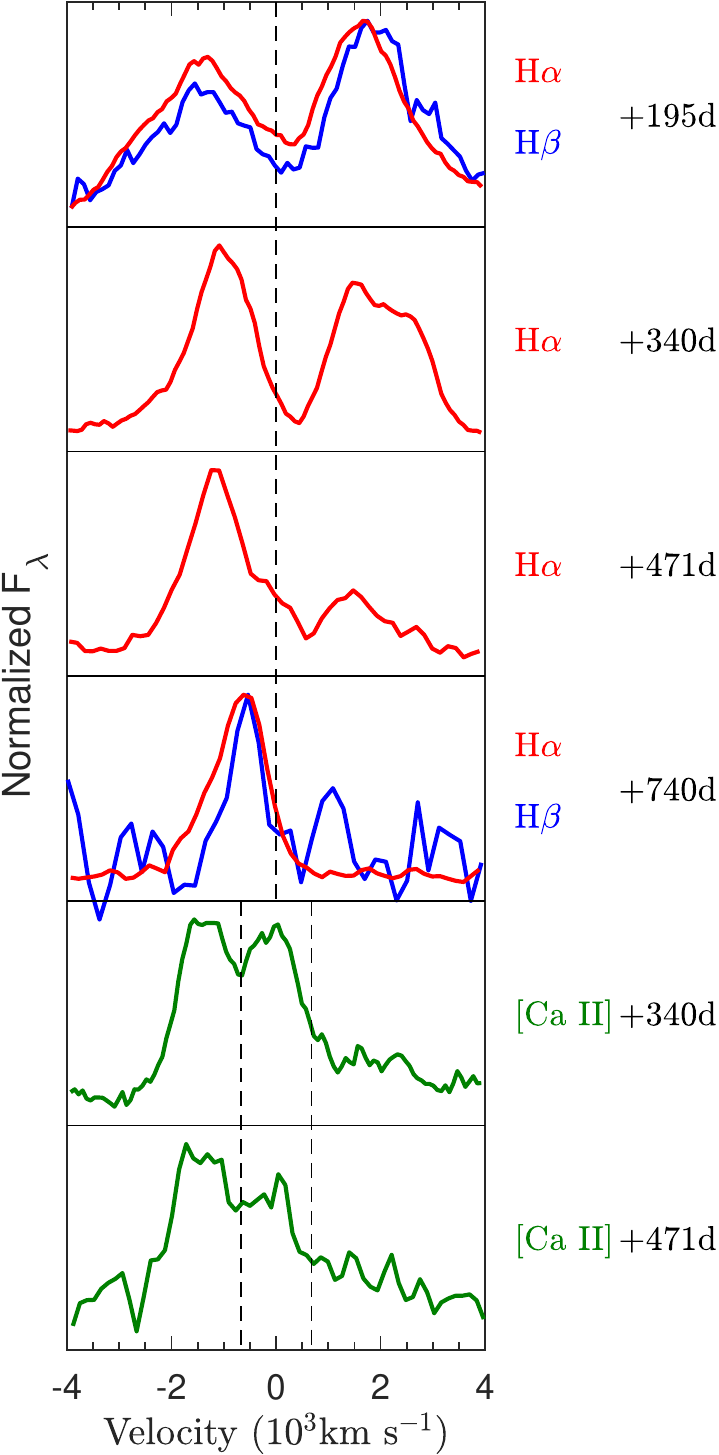}
	\caption{\ha\, \hb\ and [\Caii] nebular lines shown in the velocity domain with respect to the rest wavelengths of \ha, \hb\ and an average of [\Caii] doublet. The vertical dashed lines mark the positions of zero velocity for \ha\, \hb\ and [\Caii] doublet.}

	\label{fig:lineshift}
\end{figure}

\section{Observations} \label{sec:obsv}
\sn\ was discovered in the host galaxy \host\ by the All-Sky Automated Survey for Supernovae \citep[ASAS-SN;][]{2014ApJ...788...48S} on UT 2016-01-20.59 using the ``Brutus" telescope in Haleakala, Hawaii \citep{2016ATel.8566....1B,2017MNRAS.471.4966H}. The first ASAS-SN detection was at $V=16.81\pm0.26$\,mag on UT 2016-01-19.49, and the last non-detection was $V<18$\,mag on UT 2016-01-18.35. We adopt the explosion epoch of 2016-01-18.92 ($\rm\EpEpoch\pm0.57$) and use this as the reference epoch throughout the paper. The host galaxy distance is $ 15.2\pm3.0 $ Mpc according to \citet[Tully-Fisher distance;][]{2014MNRAS.444..527S}. We ignore any host-galaxy extinction since we do not detect any \Nai~D absorption in the SN spectrum, which is consistent with the fairly isolated location of the SN in the outskirts of the host galaxy. We adopt a total line-of-sight reddening entirely due to Milky-Way of $\ebv=0.019$ mag \citep{2011ApJ...737..103S} and $R_V=3.1$.
 
We obtained near-ultraviolet (NUV) through near-infrared (NIR) photometry and optical spectroscopy of \sn\ from 0.6d to 881d. The NUV observations were obtained with the Neil-Gehrels-Swift-Observatory UVOT. X-ray observations were obtained using the \swift\ XRT and \chandra. We summarize our optical photometric and spectroscopic observations in the Appendix, and the photometric results are reported in Table.~\ref{tab:photsn_simple}, and logs of the spectroscopic observations are given in Table~\ref{tab:speclog}.

\begin{figure*}
	\centering
	\includegraphics[width=0.75\linewidth]{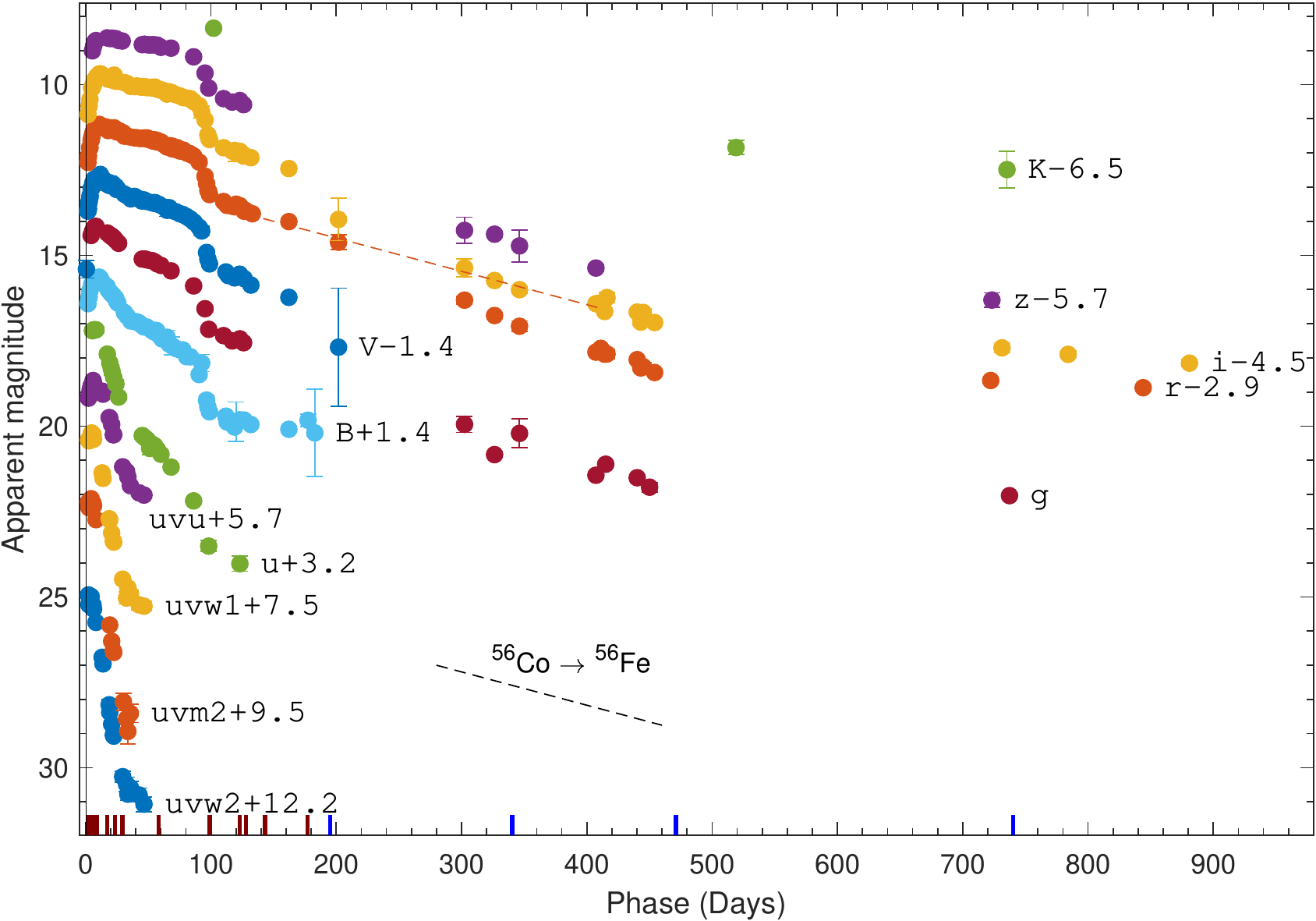}
	\caption{The photometric evolution of \sn\ in the \swift\ NUV, optical $BVugriz$-bands and NIR $K$-band. Epochs of spectral observations are marked by vertical bars at the bottom, and the four nebular spectra showing double-peaked emission are marked in blue. {Within $~200-500$ days, all the light curves declined at a faster rate than radioactive \cobalt\ decay, which is illustrated only for the \textit{r} band by a red dashed line representing \cobalt\ decay rate.}}
	\label{fig:all_lc}
\end{figure*}

\section{Results} \label{sec:results}
Optical spectra were obtained from 2d to a late nebular phase of 740\,d. We find that the early-phase ($\lesssim100$\,d) spectroscopic properties of \sn\ are typical for a normal SNe~IIP, in agreement with \citet{2018MNRAS.475.3959H}. However, the nebular spectra taken at $\gtrsim200$\,d make this SN IIP exceptional. Fig.~\ref{fig:spec_nebular} shows the nebular phase spectra of \sn\ at 195, 340, 471 and 740\,d. Starting from 195\,d, {which is $\sim100$ days after the onset of the radioactive tail phase} the \ha\ and \hb\ emissions show a very unusual double-peaked profile, where the two peaks, separated by $\sim3\times10^3$\kms, are positioned almost symmetrically in velocity with respect to the rest frame of the host galaxy. Among the two components, the relative strength of the red component is seen to be progressively decreasing with time relative to the blue component (see Fig.~\ref{fig:lineshift}). At 740d, the red component is no longer detectable, whereas the blue component of both \ha\ and \hb\ remains visibly strong, albeit narrower and shifted closer to the rest frame than in the earlier spectra. See Table~\ref{tab:profile_pars} in the Appendix for the parameters estimated from the \ha\ and \hb\ line profiles. The \ha\ emission line is detected in all the nebular spectra, while \hb\ is not clearly seen in the spectra taken at 340 and 471d, both of which have a relatively low signal-to-noise ratio.
The other nebular lines detectable in these spectra are \Nai~D (\ld5890), \Oi\ (\ld7774), \Caii\ triplets (\ldld8498, 8542, 8062) and the strong emission of [\Caii] (\ldld7291, 7324). 

The photometric light curves span from 0.6\,d to 881\,d and are shown in Fig.~\ref{fig:all_lc}. 
\citet{2018MNRAS.475.3959H} presented NUV and optical light curves up to 170d, and our photometric measurements during this period are consistent with their results.
The NUV, optical and NIR light curves show the typical evolution of SN IIP, except in the late nebular phase ($\gtrsim500$\,d), where the light curves decline more slowly than the typical SNe IIP, especially in the $g$ and $r$ bands. The flux contamination from the host galaxy may not be entirely negligible at this very late phase, but our analysis using archival Pan-STARRS1 images \citep{2016arXiv161205560C} suggests that the observed flattening is mostly intrinsic to the SN.

\sn\ was observed in 0.3-10 keV using the \swift\ and \chandra\ X-ray telescopes during $ 2-18 $~d (see Fig.~\ref{fig:xray_lc} in Appendix).
The X-ray luminosity during the initial phases of $2-5$\,d is about $25\times10^{38}$\ergs, which is around the typical luminosity for SNe II with X-ray detections \citep[see, e.g.][]{2002ApJ...572..932P, 2012MNRAS.419.1515D}. However, the luminosity declined to $5\times10^{38}$\ergs by $\sim19$\,d.

\begin{figure}
	\centering
	\includegraphics[height=14cm]{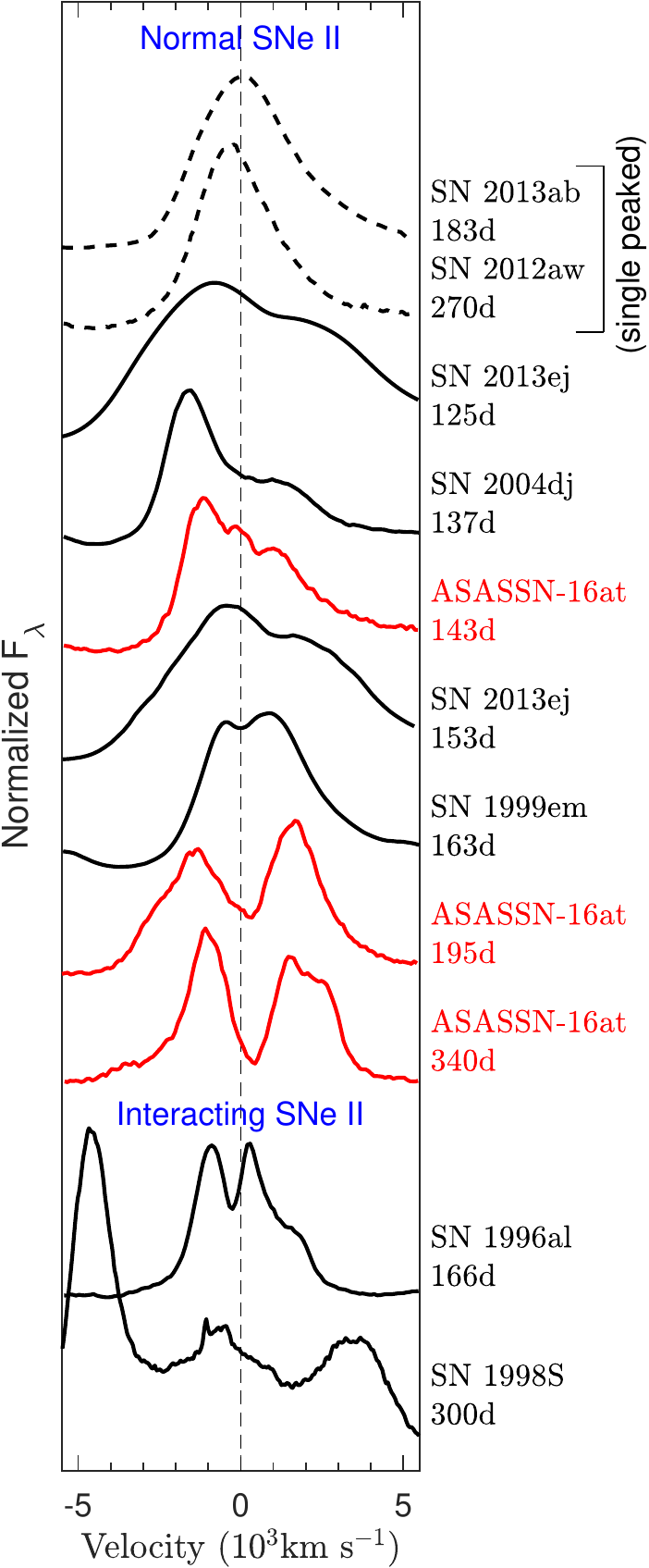}
	\caption{{The double peaked \ha\ profile of \sn\ as compared to both interacting type II SNe as well as normal SNe IIP/L having asymmetry in inner ejecta. 
	For comparison, SNe~2012aw \citep{2013MNRAS.433.1871B} and 2013ab \citep{2015MNRAS.450.2373B} are also shown as typical of the majority of normal type IIP/L SNe, which have symmetric, single component line profiles, indicating a symmetrical inner ejecta. SNe~2013ej \citep[ 153d spectrum from WISeREP]{2015ApJ...806..160B}, 2004dj \citep{2006MNRAS.369.1780V} and 1999em \citep{2002PASP..114...35L} are also normal type IIP/L SNe but show asymmetric nebular phase spectra due to asymmetry in the \nickel\ distribution or the inner line forming region. On the other hand, SNe~1996al \citep{2016MNRAS.456.3296B} and 1998S \citep{2004MNRAS.352..457P} are type~II SNe with strong ejecta-CSM interaction signatures seen in the early time, as well as nebular spectra with multi- or double- component line profiles.}
}
	\label{fig:spec_comp}
\end{figure}

\begin{figure}
	\centering
	\hspace*{0.1cm}
	\includegraphics[width=0.98\linewidth]{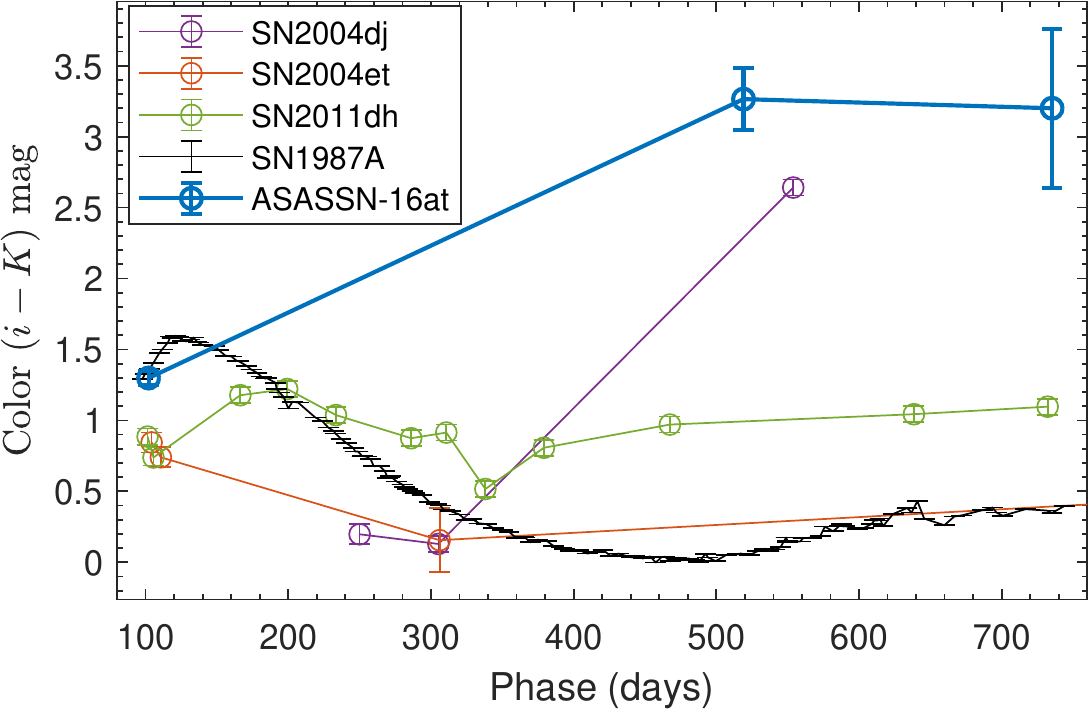}\\
	\includegraphics[width=\linewidth]{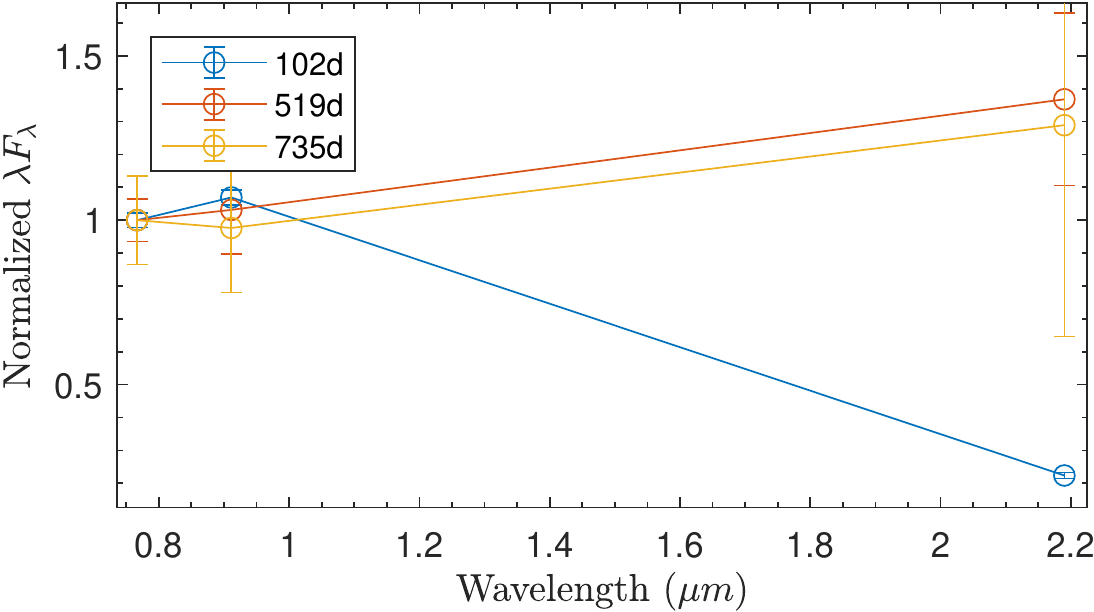}
	\caption{The top panel shows the post-photospheric phase $(i-K)$ optical-to-NIR colors of \sn\ as compared to other SNe II, which were also shown to have late-time dust formation with MIR observations. The red color of \sn\ offers supportive evidence for late-time dust formation. The bottom panel shows the evolution of the \textit{i}, \textit{z}, \textit{K} band SED normalized to the \textit{i}-band fluxes. The change of the SED shape in later phases, peaking towards IR, also supports dust formation.}
	\label{fig:nir_color}
\end{figure}

\section{Discussion} \label{sec:discussion}

{Some SNe II} shows some level of asymmetry in their nebular \ha\ emission, and a bipolar or asymmetric distribution of radioactive \nickel\ in the inner ejecta has been invoked as a possible interpretation. Some notable SNe with nebular \ha\ asymmetries include SNe~1987A \citep{1995A&A...295..129U}, 1999em \citep{2003MNRAS.338..939E}, 2004dj \citep{2005AstL...31..792C,2006AstL...32..739C}, 2012A \citep{2013MNRAS.434.1636T} and 2013ej \citep{2015ApJ...806..160B,2017MNRAS.472.5004U}. However, none of the SNe IIP observed to date have shown such a prominent bifurcation in nebular \ha\ and \hb\ emissions as in \sn, where each component of the double-peaked structure is distinctly resolved{ (see Fig.~\ref{fig:spec_comp} for comparison)}. While there might be alternative scenarios such as self-absorption, we interpret this double-peaked profile to be most likely due to strong bipolarity in the \nickel\ distribution {or in the line-forming region within the inner ejecta}. {In SNe II, the radioactive \nickel\ is likely well mixed with the ejecta \citep[e.g.,][]{1990ApJ...360..242S}, so the bipolarity seen in the inner-ejecta may imply the bipolarity in the \nickel\ distribution. Alternatively, the line-forming region is bipolar.} The prominent bifurcation of the line about the rest position likely suggests that the {the inner ejecta is composed of} a pair of almost-detached blobs. 

An additional intriguing feature of the nebular \Hi\ emission seen in \sn\ is the time evolution of its morphology. The red component of the \ha\ and \hb\ emission progressively decreases in strength relative to the blue component. This evolution can be a direct consequence of differential extinction due to late-time dust formation in the inner ejecta. The red component is emitted from regions located on the far side of the ejecta while the blue component is emitted on the near side. Consequently, the redder component suffers more line-of-sight extinction as compared to the bluer component due to the dust formed within the ejecta.  As more and more dust is formed with, the redder component suffers increasing extinction causing it to diminish in relative strength. Ultimately at 740d, the redder component is completely obscured, while the bluer component of \ha\ and \hb\, is still well detected at $\sim-600$\,\kms. {The alternative scenario explaining the blue-to-red ratio evolution of the \Hi\ profile, which we view as less likely, is that the changes are intrinsic to the emitting regions of inner ejecta. 
However, the change in shape of the red-component among +195\,d, +340\,d and +471\,d is likely intrinsic to emitting regions.
} 

The effect of differential extinction due to dust is also consistent with the blue-skewed profiles of the [\Caii] doublet (\ldld~7291, 7342) in the nebular spectra shown in Fig.~\ref{fig:lineshift}. Because the [\Caii] emission feature is intrinsically a doublet, it is not straightforward to determine if it has a double-peaked profile. However, it is clear that the peaks of the doublets are shifted blue-ward with respect to the rest frame.
{From 340d to 471d, the redder component is similarly suppressed, though to a less degree, as we could see for the \Hi\ line evolution at the same epochs. This change in line profile may further support increasing dust extinction 
if we assume the profile is dominantly the double-peaked components of [\Caii], instead of just the resolved doublet.
}
This is similar to what was observed in the [\Caii] nebular emission of SN~2007od \citep{2011MNRAS.417..261I}, which was also interpreted as a result of dust extinction in the ejecta.

To further examine dust formation, we investigate the $K$-band photometry. Ideally a quantitative comparison should be made with theoretical expectations, but robust calculations predicting NIR SED evolution for SNe II do not exist to our knowledge, thus we take an empirical approach by comparing with $K$-band observations of other SNe II. The top panel of Fig.~\ref{fig:nir_color} shows the evolution of the ($i-K$) color, which becomes significantly redder between 102\,d ($\sim1.5$\,mag)  and 735\,d ($\sim3.6$\,mag).
Since \ha\ and \hb\ line emission contribute strongly to the \textit{g} and \textit{r}-band fluxes, these bands are not used to evaluate the optical-to-NIR color or to construct the SED. 
For comparison, we also show the evolution of the $(i-K)$ color of{ SNe~1987A \citep[and references therein]{1993A&A...273..451B}, 2004dj \citep[IIP;][]{2011ApJ...732..109M}, 2004et \citep[IIP;][]{2010MNRAS.404..981M,2009ApJ...704..306K}, 2011dh \citep[IIb;][]{2015A&A...580A.142E}, which also had evidence of dust formation at late times based on their Mid-IR (MIR) observations with estimated dust masses of $ M_{dust}\approx10^{-4} - 10^{-3} \msun$ at similar phases. The optical-to-NIR color of \sn\ is significantly redder than the comparison sample at all phases and shows {an increasing trend} in reddening, supporting late-time dust formation in \sn\ and possibly more than the comparison SNe.} The SED (the bottom panel of Fig.~\ref{fig:nir_color}) evolves from a peak in optical to peak toward IR, {again supporting the} formation of dust during this period. 
{We note that the \ha\,to \hb\,line ratios are $5.7\pm0.7$ at 195\,d and $3.9\pm1.2$ at 740\,d, respectively, which is apparently at odds with the simple expectation that growing dust reddening should increase the ratio, while this comparison is subject to large uncertainty since \hb\, at 740\,d is is only detected at the $\sim 3\sigma$ level and is noise dominated.}

The optical light curve (Fig.~\ref{fig:all_lc}) followed the radioactive \cobalt\ decay during the early tail phase ($\sim100$ to $200$\,d), while thereafter ($\gtrsim200$\,d) it was dimmer than the initial \cobalt\ tail, and faded at a faster rate of  $\approx1.3$~\maghundred between $\sim300-500$\,d. Similar light-curve evolution was seen for SN~2004dj, which had strong evidence of dust formation. 
On the other hand, the flattening of the very late time ($>500$\,d) light curves of \sn\ (as mentioned in \S\ref{sec:results}) do not fit well with the proposed dust formation. Although the exact source of the flattening is unclear, it may be explained by additional flux from light echo or from the onset of a weak ejecta-CSM interaction. Such flattening at very late times was observed in SN~2007od \citep{2011MNRAS.417..261I}. Another unusual aspect of \sn\ is weak emission lines in the late-nebular spectra, especially for [\Oi] (\ldld 6300, 6364), which is one of the strongest nebular emission features for SNe with massive progenitors like SNe II. In some SNe~IIn we observe such missing nebular features as the dense CSM obscures most of the emission from the SN. This does not seem to be the case for \sn, as we do not see any evidence for dense CSM during its entire evolution, which is discussed further below. One possibility is that the dust diminishes most emission lines from the SN, while only the strongest \Hi\ and \Caii\ emissions remain detectable.

The asymmetry of \Hi\ emission lines is sometimes attributed to CSM interaction since 
CSM distributions can be sculpted to produce asymmetric line profiles as the ejecta interacts with the CSM. For instance, a triple-component profile was seen in the SN IIL/n 1996al, which was attributed to interaction with a highly asymmetric CSM \citep{2016MNRAS.456.3296B} or multiple peaks seen in the strongly interacting SN IIL/n 1998S  \citep{2004MNRAS.352..457P, 2005ApJ...622..991F}. But in these cases there is always a nebular emission component in the rest position, which is associated with the SN ejecta itself {(see Fig.~\ref{fig:spec_comp}). Interacting SNe with no asymmetry in the inner regions should always show an emission component near zero velocity, which is from the symmetric inner region. Furthermore, for the interacting SNe in our comparison sample, the central component also becomes more prominent as the SN evolves to deeper nebular phase.} Such a component at rest is absent in the nebular \Hi\ profile for \sn, thus we argue that CSM interaction alone cannot explain the observed double-peaked profile without a bipolar inner ejecta. In addition, the lack of IIn-like features in the early-phase spectra of \sn\ suggests the absence of a dense CSM, so there is no other conceivable mechanism which can obscure the \Hi\ emission from the SN itself, as it is generally discussed in cases of SNe IIn. 
Furthermore, the level of X-ray emissions from \sn\ is typical of normal SNe II \citep{2012MNRAS.419.1515D}, suggesting that the progenitor does not have a dense CSM but rather has a typical stellar wind from a RSG star with a nominal mass loss rate of $\sim 10^{-6}-10^{-7}\msun~\rm yr^{-1}$. As the shock expands, the density becomes too low to produce X-rays or to radiate the X-rays fast enough to compete with adiabatic losses and eventually the X-ray emission fades.

Among the few dozen SNe IIP with nebular phase spectra in the literature \citep[e.g.,][]{2012MNRAS.420.3451M, 2017MNRAS.467..369S}, there is no object with distinctly double-peaked nebular Balmer lines as seen in \sn. Systematic studies will be needed to establish the frequency of such profiles and their correlation with other SN properties. \sn\ demonstrates the importance of late-time observations for SNe II, even for those with rather ``normal" properties shown during the photospheric phase. 

\acknowledgments
We thank Juna A. Kollmeier who enabled our Magellan observations of this target during her program. We thank Andrea Pastorello for help in acquiring observational data and Eran Ofek and Boaz Katz for useful discussions.  
We thank Belinda Wilkes, late Neil Gehrels, and the \swift\ team for \chandra\ DDT and \swift\ ToO requests.
We acknowledge the SN spectral repository WISeREP (http://wiserep.weizmann.ac.il). 
SB, SD, and PC acknowledge NSFC Project 11573003. SB is partially supported by China postdoctoral science foundation grant No. 2018T110006. MS is supported  in part by a grant (13261) from VILLUM FONDEN. D.G. acknowledges support by SAO grant \#DD6-17079X. This paper is also partially based on observations collected at INAF Copernico 1.82m telescope; Galileo 1.22m telescope of the University of Padova;  SB, LT are partially supported by the PRIN-INAF 2016 with the project ``Toward the SKA and CTA era: discovery, localisation, and physics of transient sources''. TAT is partially supported by a Simons Foundation Fellowship and an IBM Einstein Fellowship from the Institute for Advanced Study, Princeton. N.E.R. acknowledges support from the Spanish MICINN grant ESP2017-82674-R and FEDER funds. C.G. appreciates the Carlsberg Foundation funding. MG is supported by the Polish National Science Centre grant OPUS 2015/17/B/ST9/03167. This research uses data obtained through TAP.  Partly based on observations made with GTC, installed in the Spanish Observatorio del Roque de los Muchachos of the Instituto de Astrofisica de Canarias (IAC). 
 NUTS is funded in part by the Instrument center for Danish Astrophysics (IDA). 
ASAS-SN is supported by the Gordon and Betty Moore Foundation
through grant GBMF5490 to OSU and NSF grant AST-1515927. Development of ASAS-SN has
been supported by NSF grant AST-0908816, the Mt. Cuba
Astronomical Foundation, CCAPP at OSU, CAS-SACA, the Villum Foundation, and George Skestos.
Partly based on observations made with the NOT, operated by the Nordic Optical Telescope Scientific Association at the Observatorio del Roque de los Muchachos, La Palma, Spain, of IAC.
The data presented here were obtained [in part] with ALFOSC, which is provided by the Instituto de Astrofisica de Andalucia (IAA) under a joint agreement with the University of Copenhagen and NOTSA.
%\clearpage
% Bibliography thru ms.bib; it generated ms.bbl ................. 

\pagebreak
\clearpage
\appendix

%%\section{Nebular \ha\ profile}
%%\label{app:nebular_ha}
%%{The unusual notch in nebular \ha\ profile can be described as a superimposition of two profiles. Fig.~\ref{fig:profile_combi} shows the observed \ha\ profile at 125d which is fitted by two component Gaussian profiles. These two profiles are separated by 55 \AA\ ($ \sim2500 $ \kms), one being blue shifted by -1300 \kms\ while the other is red shifted at 1200 \kms\ with respect to rest \ha\ position. The FWHM for the blue component is 54\AA\ and for red component is 146\AA. The redshifted component is dominant in strength over the blue one, having their ratio of equivalent widths to be 4.5.}
%%It may be noted that for the sake of simplicity and only for the purpose of illustration we used Gaussian profiles, which does not account for the P-Cygni absorption troughs as we see on bluer wings of line profiles in observed SN spectrum.

\begin{table}[]
	\centering
	\caption{Parameters estimated for the double-peaked profile of \ha\ and \hb\ for the nebular spectra.\label{tab:profile_pars}}
	\begin{tabular}{ll|ccccc}
\hline
Phase$ ^{a} $ & Line      & Peak Intensity Ratio & \multicolumn{2}{l}{Blue component ($10^3$ \kms)} & \multicolumn{2}{l}{Red component ($ 10^3 $ \kms)} \\
(days) &           &  (blue/red)               & Peak shift                   & FWHM                  & Peak shift                  & FWHM                  \\
\hline
		195.1 & $H\alpha$ & $0.79$                       & $-1.44$      & $2.37$ & $1.69$       & $1.83$ \\
		        & $H\beta$& $0.64$                       & $-1.46$      & $1.89$ & $1.81$       & $1.45$ \\
		340.4 & $H\alpha$ & $1.24$                       & $-1.06$      & $1.51$ & $1.65$       & $1.72$ \\
		471.1 & $H\alpha$ & $2.83$                       & $-1.14$      & $1.40$ & $1.51$       & $1.61$ \\
		740.3 & $H\alpha$ & $ \gtrsim20 $                & $-0.64$      & $1.24$ & ---        & ---  \\
		& $H\beta$  & $ \gtrsim3 $                       & $-0.56$      & $0.74$ & ---        & --- \\
		\hline
	\end{tabular}
\begin{flushleft}
	$^{a}$With reference to the explosion epoch \EpEpoch.\\
	The measured shifts and FWHMs are in units of velocity ($ 10^3 $\kms) and  are estimated by fitting two Gaussian profiles simultaneously.\\
	The parameters for \hb\ are given only when they were detectable in spectra.
\end{flushleft}
\end{table}

\section{Analysis of double-peaked line profiles} \label{sec:double_peak}
The Table~\ref{tab:profile_pars} lists the parameters estimated for each component of the double-peaked nebular emission for \ha\ and \hb.
We estimated the ratio of peak flux for the blue to the red component for each of the emissions. The FWHM and the shifts (from the rest position) for each of the red and blue components were measured by simultaneously fitting two Gaussian profiles after subtracting a local pseudo-continuum. At 340.4d the red component of the \ha\ emission is irregular in shape and does not represent a Gaussian profile, so the shift is estimated by directly measuring the maximum of the emission peak. For 740.3\,d only blue components for both \ha\ and \hb\ are visible, while red components are not detectable to the limits of signal-to-noise-ratio of the spectrum. Therefore, the ratio of peak intensity is given as a rough upper limit by considering the noise from the immediate continuum of the visible emissions.

\begin{figure*}
	\centering
	%\hspace{-0.5cm}
	\includegraphics[width=\linewidth]{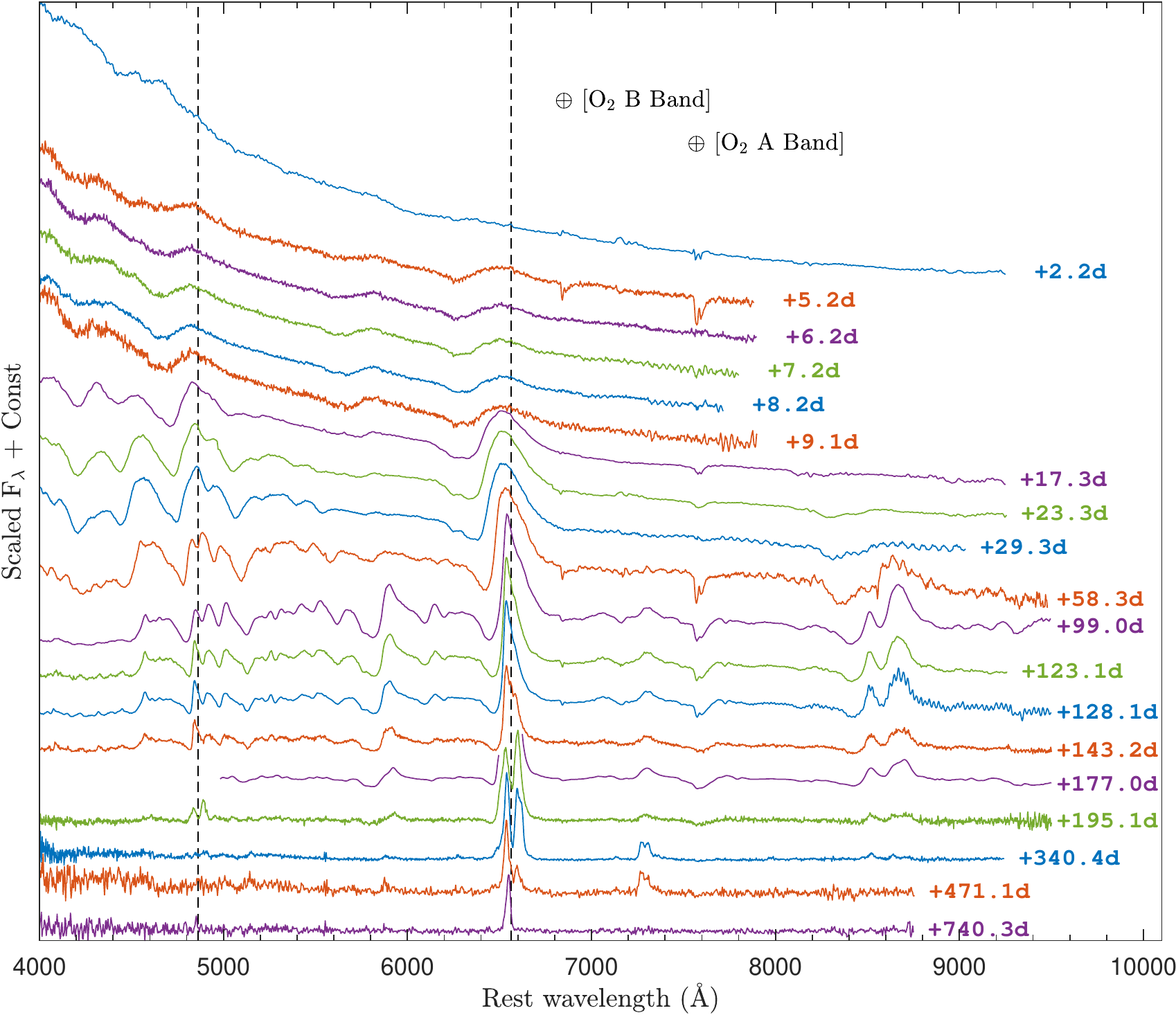}
	\caption{Spectral evolution of \sn\ from 2d to 740d. The vertical dashed lines are shown to mark the rest positions of \ha\ and \hb. The peak of the \ha\ is saturated in the spectrum observed with GTC/OSIRIS on 177.0d.}
	\label{fig:all_spec}
\end{figure*}

\section{Observing Instruments and Data}
\subsection{Photometry and Spectroscopy}

Photometric observations were obtained using the ASAS-SN quadruple 14cm ``Brutus" telescopes, the 2.0m Liverpool telescope (LT), the Las Cumbres Observatory 1.0m telescope network and the 2.6m Nordic Optical Telescope (NOT). Spectroscopic observations were done using 
the ALFOSC the 2.6m NOT,
the B\&C spectrograph on the 1.2m Galileo Telescope, the AFOSC spectrograph on the 1.8m Copernico telescope in Asiago (Italy), the SPRAT spectrograph mounted on Liverpool Telescope, the B\&C Spectrograph on the 2.5m Iren\'ee du Pont,
LDSS on the 6.5m Magellan Baade telescope, LRS on the 3.6m Telescopio Nazionale Galileo (TNG) and the OSIRIS spectrograph on 10.4m Gran Telescopio Canarias (GTC). 

Optical and near infrared photometric images were reduced using standard \iraf\ tasks and PSF photometry was performed using the \daophot\ package. The PSF radius and sky region were adjusted according to the FWHM of each image. Photometric calibrations were done using catalogs of standard stars available in the SN field. The APASS \citep[DR9;][]{2016yCat.2336....0H} catalog was used for calibrating the \textit{B} and \textit{V} band data, SDSS standards were used for the \textit{u}, \textit{g}, \textit{r}, \textit{i} and \textit{z} band data, and the 2MASS \citep{2006AJ....131.1163S} catalog was used for calibrating \textit{K}-band data. No template subtraction has been done for the optical bands, as the SN is still detectable in our latest observations. {For \textit{K} band, the host galaxy contribution is subtracted using a template observed with NOTCam at 1127\,d when the SN was no longer detectable.}
The Swift/UVOT photometry was measured with the UVOTSOURCE task in the Heasoft package using $ 5" $ apertures and placed in the Vega magnitude system, adopting the revised zero points and sensitivity from \cite{2011AIPC.1358..373B}. UVOT template images were also obtained on 2017-01-10, which are used to subtract the host contamination from SN observations.
The photometric data of \sn\ are reported in Table~\ref{tab:photsn_simple}. 

Spectroscopic data were reduced and calibrated using standard procedures of \iraf\ including cosmic-ray removals. Observations of appropriate spectrophotometric standard stars were used to flux-calibrate the spectra.
The ALFOSC and AFOSC data were reduced using ALFOSCGUI\footnote{Developed by E. Cappellaro; http://sngroup.oapd.inaf.it/foscgui.html}. The log of spectroscopic observations is given in Table~\ref{tab:speclog}. Only late-nebular spectra are shown in Fig.~\ref{fig:spec_nebular}, while the full spectral sequence is shown in Fig.~\ref{fig:all_spec}. The GTC spectrum on 2016-07-13.89 (177d) has saturation in the \ha\ region, and so the emission peak is clipped in the figure.

\begin{figure}
	\centering
	\includegraphics[width=0.5\linewidth]{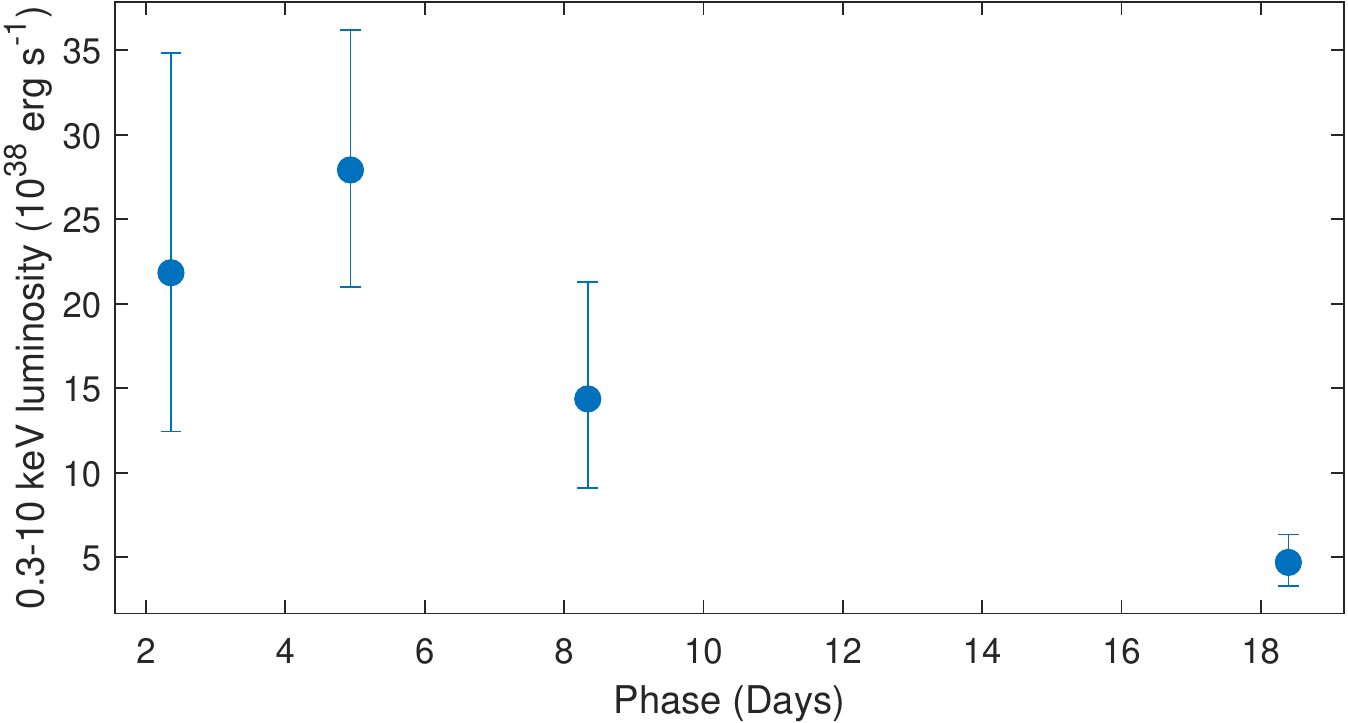}
	\caption{0.3-10 keV X-ray luminosities of \sn. The first three detections are from the \swift\ observations and while the last one is from \chandra.}
	\label{fig:xray_lc}
\end{figure}

\subsection{X-ray}
Fig.~\ref{fig:xray_lc} shows the 0.3-10 keV X-ray light curve of \sn. The first three data points were derived from the \swift\ observations while the last one was obtained from the \chandra\ observation on UT 2016-02-06. 
The \swift\ X-ray telescope (XRT)
was operating in photon counting mode and the
data were reduced by the task {\it xrtpipeline} version 0.13.1., 
which is included in the HEASOFT package 6.16. Source counts were selected in a
circle with a radius of 25$^{''}$ (10 pixels). The
background counts were collected in a nearby 
circular region with a radius of 247\farcs5. 
Due to the small number of counts used in the spectra, the counts were not binned and analyzed by applying Cash statistics.

The \swift\ fluxes were converted from count rates using {\it WPIMMS} by assuming the power law models derived from the combined \swift\ data as described below. The first data point was derived from the combined data of the first day of \swift\ observations (2016-January 21), the second from January 22 - 25, and the last \swift\ point from the observations on January-17.

After we detected \sn\ in X-rays with \swift\ \citep{2016ATel.8588....1G}, we submitted a short Director's Discretionary Time request of 5ks for \chandra\ which was approved and executed on 2016-February-06 06:38 for a total of 4963s. 
Source counts were collected in a circular region with a radius of 1$^{''}$. Background counts were collected in a nearby source-free circular region with a radius of 10$^{''}$. Source and background counts were collected in the $ 0.5-10 $ keV band. 

The supernova was clearly detected in X-rays by \chandra\ at a position of RA-2000 = 12:55:15.491\plm0.91$^{''}$ and Dec-2000 = +00:05:59.63\plm0.42$^{''}$. This position coincides with 
the optical counterpart for \sn. A total of 5 counts were detected. This results in a background corrected count rate of $1\times 10^{-3}$ counts s$^{-1}$ which is equivalent to a flux in the 0.3-10 keV band of 1.7$^{+0.5}_{-0.6} 10^{-14}~\rm erg~s^{-1}cm^{-2}$, assuming the power law spectrum with $\Gamma$=1.08 as derived from the \swift\ data (below).

\section{X-ray Spectral Analysis}
Although the number of counts from the \swift\ observations is low it still allows some limited spectral analysis. As a first step we looked at the hardness ratios that coincided with the detections shown in Figure\,\ref{fig:xray_lc}. We applied a Bayesian method
to determine the hardness ratios even with very low number statistics. The hardness ratio is defined as $HR=\frac{hard-soft}{hard+soft}$ where the soft and hard counts are in the 0.3-1.0 and 1.0-10.0 keV bands, respectively. The hardness ratios from the three detections may suggest some spectral changes. While the first data point appears to be quite hard (HR=0.72$^{+0.28}_{-0.22}$), the second data point may suggest a softening of the spectrum (HR=0.01\plm0.29) followed be a harder spectrum again (HR=0.38$^{+0.62}_{-0.24}$).

Although the supernova was faint in X-rays and the spectrum may have changed during the observation of the first week, 
the \swift\ observations of the first week (January 21-27, segments 001-012) still allow a rough spectral analysis. A total of 25 counts were collected at the source position during this time frame. 
The absorption column density was fixed to the Galactic value of $N_{\rm H}=1.40\times 10^{20}$ cm$^{-2}$ \citep{2005A&A...440..775K}. 
We first fit the spectrum with a single power law model which results in an acceptable fit (C-stat 20.3/21 degrees of freedom).
The X-ray spectral slope is very flat with a photon index $\Gamma=1.08^{+0.80}_{-0.76}$. We also fit the spectrum with a blackbody model. The blackbody  temperature equivalent energy is 1.34$^{+2.55}_{-0.49}$ keV. Although this model still resulted in an acceptable fit (27.3/21), it is less favorable than the power law model.

\begin{center}
	
{ \startlongtable
  \begin{deluxetable*}
  	{c c r c c c c c c c l}	
  	\tablecaption{Optical photometry of \sn.\label{tab:photsn_simple}}
  	
  	\tabletypesize{\fontsize{1.8mm}{2.0mm}\selectfont}
  	
\tablehead{
      UT Date &         JD $-$& Phase$^a$ &                  B &                  V &                  u &                  g &                  r &                  i &                  z & Telescope$^b$ \\
              &  2,450,000    &    (days) &              (mag) &              (mag) &              (mag) &              (mag) &              (mag) &              (mag) &              (mag) &       / Inst. }
\startdata
2016-01-19.49  &   7406.99  &     0.57     &                ---  & 16.805 $\pm$ 0.260  &                ---  &                ---  &                ---  &                ---  &                ---  &    ASASSN \\ 
2016-01-20.59  &   7408.09  &     1.67     &                ---  & 15.100 $\pm$ 0.040  &                ---  &                ---  &                ---  &                ---  &                ---  &    ASASSN \\
2016-01-20.75  &   7408.25  &     1.83     & 15.012 $\pm$ 0.044  & 15.100 $\pm$ 0.043  &                ---  &                ---  & 15.177 $\pm$ 0.018  & 15.376 $\pm$ 0.018  &                ---  &       LCOGT \\
2016-01-21.23  &   7408.73  &     2.31     &                ---  & 15.040 $\pm$ 0.050  &                ---  &                ---  & 14.992 $\pm$ 0.027  &                ---  &                ---  & ASASSN,LT \\
2016-01-21.46  &   7408.96  &     2.54     &                ---  & 14.860 $\pm$ 0.050  &                ---  &                ---  &                ---  &                ---  &                ---  &    ASASSN \\
2016-01-21.68  &   7409.18  &     2.76     & 14.820 $\pm$ 0.089  & 14.828 $\pm$ 0.069  &                ---  &                ---  &                ---  & 15.125 $\pm$ 0.049  &                ---  &       LCOGT \\
2016-01-22.16  &   7409.66  &     3.24     &                ---  &                ---  &                ---  &                ---  & 14.737 $\pm$ 0.023  &                ---  &                ---  &        LT \\
2016-01-22.32  &   7409.82  &     3.40     & 14.636 $\pm$ 0.017  & 14.703 $\pm$ 0.039  &                ---  &                ---  & 14.752 $\pm$ 0.012  & 14.933 $\pm$ 0.011  &                ---  &       LCOGT \\
2016-01-22.39  &   7409.89  &     3.47     &                ---  & 14.645 $\pm$ 0.035  &                ---  &                ---  &                ---  &                ---  &                ---  &    ASASSN \\
2016-01-23.31  &   7410.81  &     4.39     &                ---  & 14.445 $\pm$ 0.030  &                ---  & 14.416 $\pm$ 0.030  & 14.536 $\pm$ 0.020  &                ---  &                ---  & ASASSN,LT \\
2016-01-23.64  &   7411.14  &     4.72     & 14.451 $\pm$ 0.027  & 14.437 $\pm$ 0.035  &                ---  &                ---  & 14.467 $\pm$ 0.019  & 14.589 $\pm$ 0.011  &                ---  &       LCOGT \\
2016-01-24.26  &   7411.76  &     5.34     &                ---  &                ---  & 14.005 $\pm$ 0.085  & 14.306 $\pm$ 0.033  & 14.361 $\pm$ 0.023  & 14.591 $\pm$ 0.036  & 14.710 $\pm$ 0.023  &        LT \\
2016-01-24.55  &   7412.05  &     5.63     &                ---  & 14.285 $\pm$ 0.035  &                ---  &                ---  &                ---  &                ---  &                ---  &    ASASSN \\
2016-01-25.10  &   7412.60  &     6.18     &                ---  &                ---  & 13.984 $\pm$ 0.116  & 14.229 $\pm$ 0.026  & 14.270 $\pm$ 0.017  & 14.492 $\pm$ 0.036  & 14.576 $\pm$ 0.034  &        LT \\
2016-01-25.36  &   7412.86  &     6.44     &                ---  & 14.310 $\pm$ 0.030  &                ---  &                ---  &                ---  &                ---  &                ---  &    ASASSN \\
2016-01-25.62  &   7413.12  &     6.70     &                ---  & 14.170 $\pm$ 0.040  &                ---  &                ---  &                ---  &                ---  &                ---  &    ASASSN \\
2016-01-25.66  &   7413.16  &     6.74     & 14.249 $\pm$ 0.037  & 14.213 $\pm$ 0.041  &                ---  &                ---  & 14.210 $\pm$ 0.018  & 14.393 $\pm$ 0.026  &                ---  &       LCOGT \\
2016-01-26.14  &   7413.64  &     7.22     &                ---  &                ---  & 13.975 $\pm$ 0.068  & 14.164 $\pm$ 0.028  & 14.190 $\pm$ 0.022  & 14.353 $\pm$ 0.024  & 14.450 $\pm$ 0.017  &        LT \\
2016-01-26.37  &   7413.87  &     7.45     &                ---  & 14.170 $\pm$ 0.030  &                ---  &                ---  &                ---  &                ---  &                ---  &    ASASSN \\
2016-01-26.62  &   7414.12  &     7.70     &                ---  & 14.180 $\pm$ 0.030  &                ---  &                ---  &                ---  &                ---  &                ---  &    ASASSN \\
2016-01-27.07  &   7414.57  &     8.15     &                ---  &                ---  & 13.978 $\pm$ 0.094  & 14.149 $\pm$ 0.034  & 14.148 $\pm$ 0.025  & 14.311 $\pm$ 0.029  & 14.409 $\pm$ 0.016  &        LT \\
2016-01-27.62  &   7415.12  &     8.70     &                ---  & 14.170 $\pm$ 0.030  &                ---  &                ---  &                ---  &                ---  &                ---  &    ASASSN \\
2016-01-28.28  &   7415.78  &     9.36     &                ---  & 14.139 $\pm$ 0.073  &                ---  &                ---  &                ---  & 14.243 $\pm$ 0.015  &                ---  &       LCOGT \\
2016-01-28.37  &   7415.87  &     9.45     &                ---  & 14.160 $\pm$ 0.030  &                ---  &                ---  &                ---  &                ---  &                ---  &    ASASSN \\
2016-01-28.61  &   7416.11  &     9.69     &                ---  & 14.100 $\pm$ 0.040  &                ---  &                ---  &                ---  &                ---  &                ---  &    ASASSN \\
2016-01-29.23  &   7416.73  &    10.31     & 14.267 $\pm$ 0.048  & 14.234 $\pm$ 0.270  &                ---  &                ---  & 14.119 $\pm$ 0.027  &                ---  &                ---  &       LCOGT \\
2016-01-29.68  &   7417.18  &    10.76     & 14.236 $\pm$ 0.017  & 14.128 $\pm$ 0.039  &                ---  &                ---  & 14.064 $\pm$ 0.036  & 14.189 $\pm$ 0.028  &                ---  &       LCOGT \\
2016-01-30.25  &   7417.75  &    11.33     & 14.307 $\pm$ 0.019  & 14.192 $\pm$ 0.058  &                ---  &                ---  & 14.133 $\pm$ 0.022  & 14.220 $\pm$ 0.027  &                ---  &       LCOGT \\
2016-01-30.61  &   7418.11  &    11.69     &                ---  & 14.030 $\pm$ 0.030  &                ---  &                ---  &                ---  &                ---  &                ---  &    ASASSN \\
2016-01-30.65  &   7418.15  &    11.73     & 14.272 $\pm$ 0.026  & 14.146 $\pm$ 0.041  &                ---  &                ---  & 14.104 $\pm$ 0.018  & 14.182 $\pm$ 0.024  &                ---  &       LCOGT \\
2016-01-30.97  &   7418.47  &    12.05     & 14.340 $\pm$ 0.055  & 14.187 $\pm$ 0.045  &                ---  &                ---  & 14.143 $\pm$ 0.023  & 14.251 $\pm$ 0.019  &                ---  &       LCOGT \\
2016-01-31.09  &   7418.59  &    12.17     &                ---  &                ---  &                ---  &                ---  & 14.141 $\pm$ 0.039  &                ---  &                ---  &        LT \\
2016-01-31.60  &   7419.10  &    12.68     &                ---  & 14.120 $\pm$ 0.020  &                ---  &                ---  &                ---  &                ---  &                ---  &    ASASSN \\
2016-01-31.72  &   7419.22  &    12.80     & 14.357 $\pm$ 0.045  & 14.197 $\pm$ 0.043  &                ---  &                ---  & 14.126 $\pm$ 0.023  & 14.213 $\pm$ 0.016  &                ---  &       LCOGT \\
2016-02-01.47  &   7419.97  &    13.55     &                ---  & 14.150 $\pm$ 0.030  &                ---  &                ---  &                ---  &                ---  &                ---  &    ASASSN \\
2016-02-01.61  &   7420.11  &    13.69     & 14.421 $\pm$ 0.058  & 14.197 $\pm$ 0.049  &                ---  &                ---  & 14.128 $\pm$ 0.019  & 14.234 $\pm$ 0.028  &                ---  &       LCOGT \\
2016-02-03.16  &   7421.66  &    15.24     &                ---  &                ---  &                ---  &                ---  & 14.133 $\pm$ 0.018  &                ---  &                ---  &        LT \\
2016-02-03.48  &   7421.98  &    15.56     &                ---  & 14.225 $\pm$ 0.020  &                ---  &                ---  &                ---  &                ---  &                ---  &    ASASSN \\
2016-02-04.12  &   7422.62  &    16.20     &                ---  &                ---  &                ---  &                ---  & 14.148 $\pm$ 0.022  &                ---  &                ---  &        LT \\
2016-02-04.33  &   7422.83  &    16.41     &                ---  &                ---  &                ---  &                ---  &                ---  &                ---  &                ---  &    ASASSN \\
2016-02-05.03  &   7423.53  &    17.11     & 14.512 $\pm$ 0.042  & 14.273 $\pm$ 0.041  &                ---  &                ---  & 14.191 $\pm$ 0.018  & 14.287 $\pm$ 0.017  &                ---  &       LCOGT \\
2016-02-05.16  &   7423.66  &    17.24     &                ---  &                ---  & 14.689 $\pm$ 0.068  & 14.344 $\pm$ 0.024  & 14.166 $\pm$ 0.018  & 14.293 $\pm$ 0.024  & 14.331 $\pm$ 0.013  &        LT \\
2016-02-05.57  &   7424.07  &    17.65     &                ---  & 14.290 $\pm$ 0.020  &                ---  &                ---  &                ---  &                ---  &                ---  &    ASASSN \\
2016-02-05.96  &   7424.46  &    18.04     & 14.610 $\pm$ 0.056  & 14.338 $\pm$ 0.039  &                ---  &                ---  & 14.244 $\pm$ 0.019  & 14.342 $\pm$ 0.041  &                ---  &       LCOGT \\
2016-02-06.32  &   7424.82  &    18.40     &                ---  &                ---  &                ---  &                ---  &                ---  &                ---  &                ---  &    ASASSN \\
2016-02-06.60  &   7425.10  &    18.68     &                ---  & 14.270 $\pm$ 0.182  &                ---  &                ---  &                ---  &                ---  &                ---  &       LCOGT \\
2016-02-07.15  &   7425.65  &    19.23     &                ---  &                ---  & 14.942 $\pm$ 0.067  & 14.407 $\pm$ 0.026  & 14.165 $\pm$ 0.012  & 14.297 $\pm$ 0.027  & 14.349 $\pm$ 0.017  &        LT \\
2016-02-07.60  &   7426.10  &    19.68     & 14.686 $\pm$ 0.037  & 14.361 $\pm$ 0.048  &                ---  &                ---  & 14.220 $\pm$ 0.018  & 14.302 $\pm$ 0.015  &                ---  &       LCOGT \\
2016-02-08.23  &   7426.73  &    20.31     &                ---  &                ---  & 15.094 $\pm$ 0.067  & 14.420 $\pm$ 0.023  & 14.169 $\pm$ 0.018  & 14.327 $\pm$ 0.024  & 14.343 $\pm$ 0.014  &        LT \\
2016-02-08.68  &   7427.18  &    20.76     & 14.731 $\pm$ 0.046  & 14.354 $\pm$ 0.039  &                ---  &                ---  & 14.230 $\pm$ 0.022  & 14.302 $\pm$ 0.014  &                ---  &       LCOGT \\
2016-02-09.07  &   7427.57  &    21.15     &                ---  &                ---  & 15.211 $\pm$ 0.088  & 14.464 $\pm$ 0.027  & 14.181 $\pm$ 0.021  & 14.367 $\pm$ 0.025  & 14.341 $\pm$ 0.015  &        LT \\
2016-02-09.59  &   7428.09  &    21.67     & 14.759 $\pm$ 0.043  & 14.303 $\pm$ 0.053  &                ---  &                ---  & 14.226 $\pm$ 0.025  & 14.325 $\pm$ 0.023  &                ---  &       LCOGT \\
2016-02-10.25  &   7428.75  &    22.33     &                ---  &                ---  & 15.373 $\pm$ 0.066  & 14.483 $\pm$ 0.026  & 14.198 $\pm$ 0.016  & 14.336 $\pm$ 0.029  & 14.350 $\pm$ 0.018  &        LT \\
2016-02-10.62  &   7429.12  &    22.70     & 14.803 $\pm$ 0.031  & 14.376 $\pm$ 0.049  &                ---  &                ---  & 14.166 $\pm$ 0.019  & 14.222 $\pm$ 0.018  &                ---  &       LCOGT \\
2016-02-11.94  &   7430.44  &    24.02     & 14.860 $\pm$ 0.034  & 14.382 $\pm$ 0.044  &                ---  &                ---  & 14.232 $\pm$ 0.022  & 14.333 $\pm$ 0.013  &                ---  &       LCOGT \\
2016-02-12.04  &   7430.54  &    24.12     &                ---  &                ---  & 15.562 $\pm$ 0.118  & 14.560 $\pm$ 0.026  & 14.209 $\pm$ 0.016  & 14.401 $\pm$ 0.030  & 14.355 $\pm$ 0.023  &        LT \\
2016-02-12.63  &   7431.13  &    24.71     & 14.935 $\pm$ 0.038  & 14.470 $\pm$ 0.049  &                ---  &                ---  & 14.281 $\pm$ 0.024  & 14.370 $\pm$ 0.021  &                ---  &       LCOGT \\
2016-02-13.58  &   7432.08  &    25.66     & 14.996 $\pm$ 0.050  & 14.462 $\pm$ 0.037  &                ---  &                ---  & 14.288 $\pm$ 0.018  & 14.388 $\pm$ 0.015  &                ---  &       LCOGT \\
2016-02-14.19  &   7432.69  &    26.27     &                ---  &                ---  & 15.948 $\pm$ 0.075  & 14.645 $\pm$ 0.026  & 14.252 $\pm$ 0.017  & 14.402 $\pm$ 0.023  & 14.407 $\pm$ 0.023  &        LT \\
2016-02-15.06  &   7433.56  &    27.14     &                ---  &                ---  &                ---  &                ---  & 14.260 $\pm$ 0.017  &                ---  &                ---  &        LT \\
2016-02-17.05  &   7435.55  &    29.13     &                ---  &                ---  &                ---  &                ---  &                ---  &                ---  & 14.422 $\pm$ 0.016  &        LT \\
2016-02-18.07  &   7436.57  &    30.15     &                ---  &                ---  &                ---  &                ---  & 14.315 $\pm$ 0.017  &                ---  &                ---  &        LT \\
2016-02-18.76  &   7437.26  &    30.84     & 15.269 $\pm$ 0.035  & 14.606 $\pm$ 0.047  &                ---  &                ---  & 14.408 $\pm$ 0.021  & 14.442 $\pm$ 0.017  &                ---  &       LCOGT \\
2016-02-20.92  &   7439.42  &    33.00     & 15.376 $\pm$ 0.035  & 14.654 $\pm$ 0.046  &                ---  &                ---  & 14.407 $\pm$ 0.020  & 14.467 $\pm$ 0.017  &                ---  &       LCOGT \\
2016-02-23.92  &   7442.42  &    36.00     & 15.520 $\pm$ 0.037  & 14.744 $\pm$ 0.054  &                ---  &                ---  & 14.437 $\pm$ 0.019  & 14.552 $\pm$ 0.036  &                ---  &       LCOGT \\
2016-02-26.94  &   7445.44  &    39.02     & 15.538 $\pm$ 0.032  & 14.674 $\pm$ 0.042  &                ---  &                ---  & 14.458 $\pm$ 0.022  & 14.558 $\pm$ 0.022  &                ---  &       LCOGT \\
2016-02-28.64  &   7447.14  &    40.72     & 15.544 $\pm$ 0.031  & 14.714 $\pm$ 0.040  &                ---  &                ---  & 14.454 $\pm$ 0.018  & 14.538 $\pm$ 0.016  &                ---  &       LCOGT \\
2016-03-01.59  &   7449.09  &    42.67     & 15.579 $\pm$ 0.024  & 14.746 $\pm$ 0.030  &                ---  &                ---  & 14.475 $\pm$ 0.027  & 14.558 $\pm$ 0.017  &                ---  &       LCOGT \\
2016-03-03.61  &   7451.11  &    44.69     & 15.643 $\pm$ 0.031  & 14.805 $\pm$ 0.042  &                ---  &                ---  & 14.474 $\pm$ 0.020  & 14.575 $\pm$ 0.014  &                ---  &       LCOGT \\
2016-03-04.08  &   7451.58  &    45.16     &                ---  &                ---  & 17.075 $\pm$ 0.072  & 15.107 $\pm$ 0.026  & 14.466 $\pm$ 0.016  & 14.557 $\pm$ 0.025  & 14.529 $\pm$ 0.015  &        LT \\
2016-03-05.53  &   7453.03  &    46.61     & 15.694 $\pm$ 0.077  & 14.775 $\pm$ 0.041  &                ---  &                ---  &                ---  & 14.572 $\pm$ 0.020  &                ---  &       LCOGT \\
2016-03-05.99  &   7453.49  &    47.07     &                ---  &                ---  & 17.135 $\pm$ 0.072  & 15.104 $\pm$ 0.026  & 14.462 $\pm$ 0.018  & 14.558 $\pm$ 0.026  & 14.517 $\pm$ 0.014  &        LT \\
2016-03-07.67  &   7455.17  &    48.75     & 15.697 $\pm$ 0.028  & 14.821 $\pm$ 0.038  &                ---  &                ---  & 14.491 $\pm$ 0.024  & 14.588 $\pm$ 0.015  &                ---  &       LCOGT \\
2016-03-08.02  &   7455.52  &    49.10     &                ---  &                ---  & 17.195 $\pm$ 0.071  & 15.132 $\pm$ 0.025  & 14.460 $\pm$ 0.020  & 14.575 $\pm$ 0.029  & 14.532 $\pm$ 0.022  &        LT \\
2016-03-10.03  &   7457.53  &    51.11     &                ---  &                ---  & 17.452 $\pm$ 0.185  & 15.143 $\pm$ 0.025  & 14.511 $\pm$ 0.034  & 14.588 $\pm$ 0.025  & 14.538 $\pm$ 0.014  &        LT \\
2016-03-11.60  &   7459.10  &    52.68     & 15.792 $\pm$ 0.032  & 14.854 $\pm$ 0.035  &                ---  &                ---  & 14.532 $\pm$ 0.020  & 14.592 $\pm$ 0.018  &                ---  &       LCOGT \\
2016-03-11.95  &   7459.45  &    53.03     &                ---  &                ---  & 17.327 $\pm$ 0.081  & 15.154 $\pm$ 0.026  & 14.510 $\pm$ 0.025  & 14.579 $\pm$ 0.027  & 14.533 $\pm$ 0.021  &        LT \\
2016-03-13.53  &   7461.03  &    54.61     & 15.839 $\pm$ 0.049  & 14.875 $\pm$ 0.051  &                ---  &                ---  & 14.559 $\pm$ 0.028  & 14.602 $\pm$ 0.022  &                ---  &       LCOGT \\
2016-03-14.03  &   7461.53  &    55.11     &                ---  &                ---  & 17.375 $\pm$ 0.067  & 15.193 $\pm$ 0.025  & 14.516 $\pm$ 0.014  & 14.594 $\pm$ 0.025  & 14.535 $\pm$ 0.014  &        LT \\
2016-03-14.11  &   7461.61  &    55.19     & 15.796 $\pm$ 0.064  & 14.853 $\pm$ 0.037  &                ---  &                ---  & 14.537 $\pm$ 0.023  & 14.572 $\pm$ 0.016  &                ---  &       LCOGT \\
2016-03-15.07  &   7462.57  &    56.15     &                ---  &                ---  &                ---  &                ---  & 14.562 $\pm$ 0.026  &                ---  &                ---  &        LT \\
2016-03-15.51  &   7463.01  &    56.59     & 15.801 $\pm$ 0.065  & 14.884 $\pm$ 0.046  &                ---  &                ---  & 14.569 $\pm$ 0.023  & 14.621 $\pm$ 0.020  &                ---  &       LCOGT \\
2016-03-16.01  &   7463.51  &    57.09     &                ---  &                ---  & 17.478 $\pm$ 0.067  & 15.247 $\pm$ 0.025  & 14.533 $\pm$ 0.012  & 14.615 $\pm$ 0.025  & 14.543 $\pm$ 0.015  &        LT \\
2016-03-17.58  &   7465.08  &    58.66     & 15.933 $\pm$ 0.034  & 14.911 $\pm$ 0.047  &                ---  &                ---  & 14.585 $\pm$ 0.015  & 14.648 $\pm$ 0.017  &                ---  &       LCOGT \\
2016-03-19.00  &   7466.50  &    60.08     &                ---  &                ---  & 17.623 $\pm$ 0.089  & 15.294 $\pm$ 0.026  & 14.579 $\pm$ 0.013  & 14.660 $\pm$ 0.024  & 14.608 $\pm$ 0.019  &        LT \\
2016-03-19.85  &   7467.35  &    60.93     & 16.045 $\pm$ 0.060  &                ---  &                ---  &                ---  &                ---   & 14.660 $\pm$ 0.033  &                ---  &       LCOGT \\
2016-03-21.51  &   7469.01  &    62.59     &                ---  & 14.988 $\pm$ 0.055  &                ---  &                ---  & 14.661 $\pm$ 0.035  & 14.684 $\pm$ 0.032  &                ---  &       LCOGT \\
2016-03-23.84  &   7471.34  &    64.92     & 15.997 $\pm$ 0.205  & 15.078 $\pm$ 0.182  &                ---  &                ---  & 14.725 $\pm$ 0.083  & 14.777 $\pm$ 0.119  &                ---  &       LCOGT \\
2016-03-25.55  &   7473.05  &    66.63     & 16.196 $\pm$ 0.115  & 15.003 $\pm$ 0.079  &                ---  &                ---  & 14.690 $\pm$ 0.023  & 14.710 $\pm$ 0.041  &                ---  &       LCOGT \\
2016-03-27.07  &   7474.57  &    68.15     &                ---  &                ---  & 17.998 $\pm$ 0.069  & 15.454 $\pm$ 0.025  & 14.694 $\pm$ 0.014  & 14.751 $\pm$ 0.023  & 14.632 $\pm$ 0.017  &        LT \\
2016-03-27.85  &   7475.35  &    68.93     & 16.249 $\pm$ 0.269  & 15.084 $\pm$ 0.065  &                ---  &                ---  & 14.779 $\pm$ 0.051  & 14.738 $\pm$ 0.037  &                ---  &       LCOGT \\
2016-03-28.07  &   7475.57  &    69.15     &                ---  &                ---  &                ---  &                ---  & 14.707 $\pm$ 0.015  &                ---  &                ---  &        LT \\
2016-03-28.97  &   7476.47  &    70.05     &                ---  &                ---  &                ---  &                ---  & 14.720 $\pm$ 0.020  &                ---  &                ---  &        LT \\
2016-03-30.85  &   7478.35  &    71.93     & 16.326 $\pm$ 0.044  & 15.121 $\pm$ 0.047  &                ---  &                ---  & 14.800 $\pm$ 0.018  & 14.797 $\pm$ 0.025  &                ---  &       LCOGT \\
2016-03-31.96  &   7479.46  &    73.04     &                ---  &                ---  &                ---  &                ---  & 14.746 $\pm$ 0.023  &                ---  &                ---  &        LT \\
2016-04-03.08  &   7481.58  &    75.16     & 16.366 $\pm$ 0.041  & 15.180 $\pm$ 0.050  &                ---  &                ---  & 14.843 $\pm$ 0.017  & 14.829 $\pm$ 0.024  &                ---  &       LCOGT \\
2016-04-05.50  &   7484.00  &    77.58     & 16.369 $\pm$ 0.037  & 15.203 $\pm$ 0.048  &                ---  &                ---  & 14.843 $\pm$ 0.018  & 14.877 $\pm$ 0.022  &                ---  &       LCOGT \\
2016-04-05.98  &   7484.48  &    78.06     &                ---  &                ---  &                ---  &                ---  & 14.818 $\pm$ 0.024  &                ---  &                ---  &        LT \\
2016-04-08.54  &   7487.04  &    80.62     & 16.565 $\pm$ 0.060  & 15.266 $\pm$ 0.040  &                ---  &                ---  & 14.900 $\pm$ 0.021  & 14.894 $\pm$ 0.023  &                ---  &       LCOGT \\
2016-04-10.95  &   7489.45  &    83.03     &                ---  &                ---  &                ---  &                ---  & 14.919 $\pm$ 0.016  &                ---  &                ---  &        LT \\
2016-04-11.50  &   7490.00  &    83.58     & 16.569 $\pm$ 0.047  & 15.335 $\pm$ 0.049  &                ---  &                ---  & 14.941 $\pm$ 0.024  & 14.912 $\pm$ 0.027  &                ---  &       LCOGT \\
2016-04-14.08  &   7492.58  &    86.16     &                ---  &                ---  & 18.988 $\pm$ 0.073  & 15.893 $\pm$ 0.025  & 14.989 $\pm$ 0.018  & 14.997 $\pm$ 0.026  & 14.886 $\pm$ 0.014  &        LT \\
2016-04-14.50  &   7493.00  &    86.58     &                ---  & 15.428 $\pm$ 0.039  &                ---  &                ---  &                ---  &                ---  &                ---  &       LCOGT \\
2016-04-18.47  &   7496.97  &    90.55     & 17.089 $\pm$ 0.071  & 15.577 $\pm$ 0.045  &                ---  &                ---  & 15.168 $\pm$ 0.028  & 15.117 $\pm$ 0.021  &                ---  &       LCOGT \\
2016-04-20.51  &   7499.01  &    92.59     & 16.753 $\pm$ 0.252  & 15.684 $\pm$ 0.147  &                ---  &                ---  &                ---  & 15.302 $\pm$ 0.169  &                ---  &       LCOGT \\
2016-04-23.16  &   7501.66  &    95.24     &                ---  &                ---  &                ---  & 16.561 $\pm$ 0.039  & 15.584 $\pm$ 0.039  & 15.532 $\pm$ 0.028  & 15.362 $\pm$ 0.025  &        LT \\
2016-04-24.39  &   7502.89  &    96.47     & 17.837 $\pm$ 0.144  & 16.315 $\pm$ 0.052  &                ---  &                ---  & 15.798 $\pm$ 0.043  &                ---  &                ---  &       LCOGT \\
2016-04-25.43  &   7503.93  &    97.51     & 18.042 $\pm$ 0.067  & 16.506 $\pm$ 0.044  &                ---  &                ---  & 15.996 $\pm$ 0.021  & 15.968 $\pm$ 0.018  &                ---  &       LCOGT \\
2016-04-26.10  &   7504.60  &    98.18     &                ---  &                ---  & 20.313 $\pm$ 0.159  & 17.164 $\pm$ 0.028  & 16.026 $\pm$ 0.017  & 16.062 $\pm$ 0.027  & 15.800 $\pm$ 0.019  &        LT \\
2016-04-26.75  &   7505.25  &    98.83     & 18.180 $\pm$ 0.068  & 16.649 $\pm$ 0.031  &                ---  &                ---  & 16.119 $\pm$ 0.014  & 16.108 $\pm$ 0.022  &                ---  &       LCOGT \\
2016-05-07.96  &   7516.46  &   110.04     &                ---  &                ---  &                ---  & 17.349 $\pm$ 0.038  & 16.324 $\pm$ 0.031  & 16.347 $\pm$ 0.029  & 16.110 $\pm$ 0.029  &        LT \\
2016-05-09.85  &   7518.35  &   111.93     & 18.309 $\pm$ 0.067  & 16.885 $\pm$ 0.048  &                ---  &                ---  &                ---  &                ---  &                ---  &       LCOGT \\
2016-05-10.78  &   7519.28  &   112.86     & 18.477 $\pm$ 0.039  & 16.924 $\pm$ 0.034  &                ---  &                ---  & 16.436 $\pm$ 0.021  &                ---  &                ---  &       LCOGT \\
2016-05-11.85  &   7520.35  &   113.93     & 18.438 $\pm$ 0.046  & 16.992 $\pm$ 0.042  &                ---  &                ---  &                ---  &                ---  &                ---  &       LCOGT \\
2016-05-14.91  &   7523.41  &   116.99     &                ---  &                ---  &                ---  & 17.504 $\pm$ 0.028  & 16.457 $\pm$ 0.017  & 16.447 $\pm$ 0.024  & 16.211 $\pm$ 0.016  &        LT \\
2016-05-16.76  &   7525.26  &   118.84     & 18.626 $\pm$ 0.105  & 17.058 $\pm$ 0.058  &                ---  &                ---  & 16.477 $\pm$ 0.021  & 16.427 $\pm$ 0.037  &                ---  &       LCOGT \\
2016-05-17.98  &   7526.48  &   120.06     & 18.468 $\pm$ 0.574  &                ---  &                ---  &                ---  & 16.401 $\pm$ 0.103  & 16.506 $\pm$ 0.243  &                ---  &       LCOGT \\
2016-05-20.78  &   7529.28  &   122.86     & 18.414 $\pm$ 0.068  & 16.951 $\pm$ 0.050  &                ---  &                ---  & 16.452 $\pm$ 0.015  & 16.452 $\pm$ 0.037  &                ---  &       LCOGT \\
2016-05-20.99  &   7529.49  &   123.07     &                ---  &                ---  & 20.829 $\pm$ 0.221  & 17.443 $\pm$ 0.030  & 16.447 $\pm$ 0.013  & 16.460 $\pm$ 0.025  & 16.167 $\pm$ 0.014  &        LT \\
2016-05-23.94  &   7532.44  &   126.02     &                ---  &                ---  &                ---  & 17.557 $\pm$ 0.027  & 16.558 $\pm$ 0.018  & 16.593 $\pm$ 0.026  & 16.286 $\pm$ 0.017  &        LT \\
2016-05-24.38  &   7532.88  &   126.46     & 18.422 $\pm$ 0.080  & 17.079 $\pm$ 0.043  &                ---  &                ---  & 16.607 $\pm$ 0.021  & 16.564 $\pm$ 0.027  &                ---  &       LCOGT \\
2016-05-25.88  &   7534.38  &   127.96     &                ---  &                ---  &                ---  &                ---  & 16.594 $\pm$ 0.018  &                ---  &                ---  &        LT \\
2016-05-29.77  &   7538.27  &   131.85     & 18.551 $\pm$ 0.053  & 17.269 $\pm$ 0.044  &                ---  &                ---  &                ---  & 16.639 $\pm$ 0.026  &                ---  &       LCOGT \\
2016-05-30.89  &   7539.39  &   132.97     &                ---  &                ---  &                ---  &                ---  & 16.680 $\pm$ 0.017  &                ---  &                ---  &        LT \\
2016-06-29.09  &   7568.59  &   162.17     & 18.692 $\pm$ 0.041  & 17.628 $\pm$ 0.035  &                ---  &                ---  & 16.911 $\pm$ 0.017  & 16.958 $\pm$ 0.021  &                ---  &       LCOGT \\
2016-07-14.41  &   7583.91  &   177.49     & 18.427 $\pm$ 0.177  &                ---  &                ---  &                ---  &                ---  &                ---  &                ---  &       LCOGT \\
2016-07-19.80  &   7589.30  &   182.88     & 18.797 $\pm$ 1.278  &                ---  &                ---  &                ---  &                ---  &                ---  &                ---  &       LCOGT \\
2016-08-07.72  &   7608.22  &   201.80     &                ---  & 19.087 $\pm$ 1.727  &                ---  &                ---  & 17.515 $\pm$ 0.221  & 18.447 $\pm$ 0.622  &                ---  &       LCOGT \\
2016-11-16.27  &   7708.77  &   302.35     &                ---  &                ---  &                ---  & 19.947 $\pm$ 0.239  & 19.210 $\pm$ 0.112  & 19.865 $\pm$ 0.265  & 19.967 $\pm$ 0.384  &        LT \\
2016-12-10.26  &   7732.76  &   326.34     &                ---  &                ---  &                ---  & 20.833 $\pm$ 0.040  & 19.662 $\pm$ 0.028  & 20.235 $\pm$ 0.039  & 20.079 $\pm$ 0.061  &        LT \\
2016-12-30.17  &   7752.67  &   346.25     &                ---  &                ---  &                ---  & 20.211 $\pm$ 0.420  & 19.974 $\pm$ 0.157  & 20.504 $\pm$ 0.108  & 20.421 $\pm$ 0.469  &        LT \\
2017-03-01.15  &   7813.65  &   407.23     &                ---  &                ---  &                ---  & 21.436 $\pm$ 0.044  & 20.732 $\pm$ 0.036  & 20.915 $\pm$ 0.072  & 21.071 $\pm$ 0.101  &        LT \\
2017-03-05.04  &   7817.54  &   411.12     &                ---  &                ---  &                ---  &                ---  & 20.622 $\pm$ 0.050  & 20.893 $\pm$ 0.060  &                ---  &        LT \\
2017-03-08.11  &   7820.61  &   414.19     &                ---  &                ---  &                ---  &                ---  & 20.793 $\pm$ 0.051  & 21.143 $\pm$ 0.069  &                ---  &        LT \\
2017-03-08.97  &   7821.47  &   415.05     &                ---  &                ---  &                ---  & 21.114 $\pm$ 0.096  &                ---  &                ---  &                ---  &        LT \\
2017-03-10.00  &   7822.50  &   416.08     &                ---  &                ---  &                ---  &                ---  & 20.785 $\pm$ 0.139  & 20.734 $\pm$ 0.143  &                ---  &        LT \\
2017-04-03.08  &   7846.58  &   440.16     &                ---  &                ---  &                ---  & 21.510 $\pm$ 0.052  & 20.953 $\pm$ 0.046  & 21.160 $\pm$ 0.065  &                ---  &        LT \\
2017-04-05.95  &   7849.45  &   443.03     &                ---  &                ---  &                ---  &                ---  & 21.195 $\pm$ 0.061  & 21.453 $\pm$ 0.073  &                ---  &        LT \\
2017-04-07.93  &   7851.43  &   445.01     &                ---  &                ---  &                ---  &                ---  & 21.161 $\pm$ 0.120  & 21.171 $\pm$ 0.104  &                ---  &        LT \\
2017-04-12.99  &   7856.49  &   450.07     &                ---  &                ---  &                ---  & 21.787 $\pm$ 0.135  &                ---  &                ---  &                ---  &        LT \\
2017-04-16.99  &   7860.49  &   454.07     &                ---  &                ---  &                ---  &                ---  & 21.330 $\pm$ 0.040  & 21.465 $\pm$ 0.046  &                ---  &        LT \\
2018-01-10.25  &   8128.75  &   722.33     &                ---  &                ---  &                ---  &                ---  & 21.563 $\pm$ 0.067  &                ---  &                ---  &        LT \\
2018-01-11.22  &   8129.72  &   723.30     &                ---  &                ---  &                ---  &                ---  &                ---  &                ---  & 22.008 $\pm$ 0.214  &        LT \\
2018-01-19.28  &   8137.78  &   731.36     &                ---  &                ---  &                ---  &                ---  &                ---  & 22.206 $\pm$ 0.150  &                ---  &        LT \\
2018-01-25.24  &   8143.74  &   737.32     &                ---  &                ---  &                ---  & 22.034 $\pm$ 0.047  &                ---  &                ---  &                ---  &        LT \\
2018-03-13.06  &   8190.56  &   784.14     &                ---  &                ---  &                ---  &                ---  &                ---  & 22.397 $\pm$ 0.094  &                ---  &        LT \\
2018-05-11.90  &   8250.40  &   843.98     &                ---  &                ---  &                ---  &                ---  & 21.774 $\pm$ 0.038  &                ---  &                ---  &        LT \\
2018-06-17.89  &   8287.39  &   880.97     &                ---  &                ---  &                ---  &                ---  &                ---  & 22.655 $\pm$ 0.113  &                ---  &        LT \\
\enddata
\end{deluxetable*}

\addtocounter{table}{-1}
\begin{deluxetable*}
	{c c r c l}	
\tablecaption{{\it(continued)} NIR photometry.\label{tab:photsnNIR_simple}}
%\tabletypesize{\fontsize{1.8mm}{2.0mm}\selectfont}

\tablehead{
	UT Date &         JD $-$& Phase$^a$ &                  K  &Telescope$^b$ \\
	&  2,450,000    &    (days)         &              (mag)  &      / Inst. }
\startdata
%2016-01-21.49   &  7408.99  &     2.57   &   13.200 $\pm$ 0.200  &  UKIRT     \\ 
2016-04-30.02   &  7508.52  &   102.10   &   14.85 $\pm$ 0.05  &  NC     \\ 
2017-06-20.97   &  7925.47  &   519.05   &   18.34 $\pm$ 0.21  &  NC     \\ 
2018-01-23.23   &  8141.73  &   735.31   &   18.99 $\pm$ 0.54  &  NC     \\ 
2019-02-19.12   &  8533.62  &   1127.20   &   ---  &  NC     \\ 
\enddata
\end{deluxetable*}

\addtocounter{table}{-1}
\begin{deluxetable*}
	{c c r c c c c c c l}
	\caption{{\it(continued)} NUV photometry.\label{tab:photsnNUV_simple}}
\tablehead{
	UT Date &         JD $-$& Phase$^a$ &              uvw2 &              uvm2 &               uvw1 &             uvu &               uvb &             uvv & Telescope$^b$ \\
	&  2,450,000    &    (days)         &              (mag) &            (mag) &              (mag) &           (mag) &             (mag) &           (mag) &       / Inst. }
\startdata
2016-01-21.08   &  7408.58   &    2.16    &  12.742 $\pm$ 0.038 &  12.718 $\pm$ 0.039 &  12.899 $\pm$ 0.037 &  13.479 $\pm$ 0.033 &  14.829 $\pm$ 0.033 &  14.822 $\pm$ 0.045 &   UVOT   \\
2016-01-21.51   &  7409.01   &    2.59    &  13.030 $\pm$ 0.038 &  12.886 $\pm$ 0.038 &  12.930 $\pm$ 0.038 &  13.413 $\pm$ 0.035 &  14.741 $\pm$ 0.035 &  14.803 $\pm$ 0.045 &   UVOT   \\
2016-01-22.45   &  7409.95   &    3.53    &  12.891 $\pm$ 0.038 &  12.745 $\pm$ 0.039 &  12.871 $\pm$ 0.038 &  13.254 $\pm$ 0.035 &  14.550 $\pm$ 0.035 &  14.673 $\pm$ 0.045 &   UVOT   \\
2016-01-22.79   &  7410.29   &    3.87    &  12.815 $\pm$ 0.039 &  12.686 $\pm$ 0.038 &  12.826 $\pm$ 0.038 &  13.202 $\pm$ 0.034 &  14.496 $\pm$ 0.033 &  14.516 $\pm$ 0.039 &   UVOT   \\
2016-01-23.19   &  7410.69   &    4.27    &  12.789 $\pm$ 0.039 &  12.625 $\pm$ 0.039 &  12.730 $\pm$ 0.039 &  13.118 $\pm$ 0.038 &  14.366 $\pm$ 0.038 &  14.474 $\pm$ 0.051 &   UVOT   \\
2016-01-23.66   &  7411.16   &    4.74    &                 --- &                 --- &  12.700 $\pm$ 0.046 &                 --- &                 --- &                 --- &   UVOT   \\
2016-01-24.79   &  7412.29   &    5.87    &  13.037 $\pm$ 0.037 &  12.757 $\pm$ 0.038 &  12.743 $\pm$ 0.037 &  12.964 $\pm$ 0.033 &  14.234 $\pm$ 0.031 &  14.217 $\pm$ 0.032 &   UVOT   \\
2016-01-25.26   &  7412.76   &    6.34    &  13.156 $\pm$ 0.040 &  12.835 $\pm$ 0.040 &  12.884 $\pm$ 0.041 &  13.032 $\pm$ 0.041 &  14.247 $\pm$ 0.042 &  14.184 $\pm$ 0.054 &   UVOT   \\
2016-01-27.17   &  7414.67   &    8.25    &  13.546 $\pm$ 0.039 &  13.237 $\pm$ 0.039 &                 --- &                 --- &                 --- &                 --- &   UVOT   \\
2016-02-01.03   &  7419.53   &   13.11    &  14.568 $\pm$ 0.041 &                 --- &  13.868 $\pm$ 0.040 &  13.275 $\pm$ 0.035 &  14.293 $\pm$ 0.033 &  14.133 $\pm$ 0.038 &   UVOT   \\
2016-02-01.76   &  7420.26   &   13.84    &  14.760 $\pm$ 0.041 &                 --- &  14.029 $\pm$ 0.040 &  13.363 $\pm$ 0.035 &  14.298 $\pm$ 0.033 &  14.178 $\pm$ 0.038 &   UVOT   \\
2016-02-06.69   &  7425.19   &   18.77    &  15.959 $\pm$ 0.159 &                 --- &  15.209 $\pm$ 0.056 &  14.049 $\pm$ 0.043 &  14.488 $\pm$ 0.040 &                 --- &   UVOT   \\
2016-02-07.05   &  7425.55   &   19.13    &  16.177 $\pm$ 0.053 &  16.323 $\pm$ 0.059 &  15.244 $\pm$ 0.046 &  14.058 $\pm$ 0.036 &  14.535 $\pm$ 0.033 &  14.287 $\pm$ 0.039 &   UVOT   \\
2016-02-08.60   &  7427.10   &   20.68    &  16.527 $\pm$ 0.069 &  16.797 $\pm$ 0.086 &  15.627 $\pm$ 0.058 &  14.251 $\pm$ 0.042 &  14.691 $\pm$ 0.038 &  14.310 $\pm$ 0.043 &   UVOT   \\
2016-02-09.93   &  7428.43   &   22.01    &  16.841 $\pm$ 0.065 &  17.076 $\pm$ 0.077 &  15.858 $\pm$ 0.054 &  14.553 $\pm$ 0.039 &  14.683 $\pm$ 0.034 &  14.275 $\pm$ 0.036 &   UVOT   \\
2016-02-10.39   &  7428.89   &   22.47    &  16.876 $\pm$ 0.081 &  17.125 $\pm$ 0.094 &  15.894 $\pm$ 0.065 &  14.547 $\pm$ 0.045 &  14.718 $\pm$ 0.039 &  14.356 $\pm$ 0.045 &   UVOT   \\
2016-02-17.44   &  7435.94   &   29.52    &  18.064 $\pm$ 0.164 &                 --- &  16.981 $\pm$ 0.090 &  15.491 $\pm$ 0.052 &  15.151 $\pm$ 0.038 &                 --- &   UVOT   \\
2016-02-17.92   &  7436.42   &   30.00    &                 --- &  18.569 $\pm$ 0.249 &                 --- &                 --- &                 --- &  14.481 $\pm$ 0.052 &   UVOT   \\
2016-02-20.56   &  7439.06   &   32.64    &  18.286 $\pm$ 0.209 &  19.070 $\pm$ 0.363 &  17.533 $\pm$ 0.172 &  15.628 $\pm$ 0.075 &  15.228 $\pm$ 0.050 &  14.615 $\pm$ 0.059 &   UVOT   \\
2016-02-21.57   &  7440.07   &   33.65    &  18.575 $\pm$ 0.170 &  19.438 $\pm$ 0.370 &  17.232 $\pm$ 0.094 &  15.785 $\pm$ 0.052 &  15.248 $\pm$ 0.036 &  14.580 $\pm$ 0.038 &   UVOT   \\
2016-02-23.58   &  7442.08   &   35.66    &  18.387 $\pm$ 0.181 &  18.910 $\pm$ 0.268 &  17.416 $\pm$ 0.105 &  16.042 $\pm$ 0.059 &  15.348 $\pm$ 0.037 &  14.652 $\pm$ 0.049 &   UVOT   \\
2016-03-01.60   &  7449.10   &   42.68    &  18.601 $\pm$ 0.205 &                 --- &  17.733 $\pm$ 0.152 &  16.253 $\pm$ 0.077 &  15.518 $\pm$ 0.044 &  14.746 $\pm$ 0.048 &   UVOT   \\
2016-03-05.33   &  7452.83   &   46.41    &  18.868 $\pm$ 0.218 &                 --- &  17.771 $\pm$ 0.136 &  16.318 $\pm$ 0.069 &  15.607 $\pm$ 0.041 &  14.762 $\pm$ 0.042 &   UVOT   \\
\enddata

\tablecomments{\\
	$^{a}$With reference to the explosion epoch \EpEpoch.\\
$^{b}$ The abbreviations of telescope/instrument used are as follows: ASASSN - ASAS-SN quadruple 14-cm telescopes; LCOGT - Las Cumbres Observatory 1 m telescope
	network; LT - 2m Liverpool Telescope; NC - NOTCam mounted on 2.0m NOT; 
	%UKIRT - 3.8m UKIRT infrared telescope; 
	UVOT - \swift\ Ultraviolet Optical Telescope.\\
	%$^{c}$ FWHM of the median stellar PSF at $V$ band frame.\\
	Data observed within 5\,hr are represented under a single-epoch observation.}
\end{deluxetable*}
}

%\begin{flushleft}
%  $^{a}$ with reference to the explosion epoch \EpEpoch\\
%  $^{b}$ ST : 104-cm Sampurnanand Telescope, ARIES, India; DFOT : 130-cm Devasthal fast optical telescope, ARIES, India; HCT: 2m Himalyan Chandra Telescope, Hanle, India; Swift: \textit{Swift}~UVOT\\
%  %$^{c}$ FWHM of the median stellar PSF at $V$ band frame.\\
%  Note: Data observed within 5 Hrs, are represented under single epoch observation.
%\end{flushleft}
\end{center}

%\clearpage
%\end{turnpage}
% tab:speclog
% tab:speclog
%____________________________________________________________________________

\begin{table}
\centering
  \caption{Summary of spectroscopic observations of \sn.}
  \label{tab:speclog}
  \begin{tabular}{lc r l}
    \hline
    UT Date       &JD      &Phase$^{a}$&Telescope \\
                &2450000+&(days)     &/ Instrument                \\ \hline
2016-01-21.11 & 7408.61  &   2.2  & Copernico/AFOSC  \\
2016-01-24.14 & 7411.64  &   5.2  & Galileo/B\&C   \\
2016-01-25.16 & 7412.66  &   6.2  & Galileo/B\&C   \\
2016-01-26.13 & 7413.63  &   7.2  & Galileo/B\&C   \\
2016-01-27.08 & 7414.58  &   8.2  & Galileo/B\&C   \\
2016-01-28.06 & 7415.56  &   9.1  & Galileo/B\&C   \\
2016-02-05.24 & 7423.74  &  17.3  & Copernico/AFOSC  \\
2016-02-11.17 & 7429.67  &  23.3  & NOT/ALFOSC    \\
2016-02-17.20 & 7435.70  &  29.3  & NOT/ALFOSC    \\
2016-03-17.20 & 7464.70  &  58.3  & DuPont/B\&C   \\
2016-04-26.92 & 7505.42  &  99.0  & NOT/ALFOSC    \\
2016-05-20.99 & 7529.49  & 123.1  & NOT/ALFOSC    \\
2016-05-26.00 & 7534.50  & 128.1  & TNG/LRS       \\
2016-06-10.13 & 7549.63  & 143.2  & DuPont/B\&C   \\
2016-07-13.89 & 7583.39  & 177.0  & GTC/OSIRIS    \\
2016-08-01.01 & 7601.51  & 195.1  & DuPont/B\&C   \\
2016-12-24.34 & 7746.84  & 340.4  & Magellan/LDSS3\\
2017-05-03.98 & 7877.48  & 471.1  & GTC/OSIRIS    \\
2018-01-28.17 & 8146.67  & 740.3  & GTC/OSIRIS    \\
    \hline
  \end{tabular}
\begin{flushleft}
  $^{a}$ The phase is the number of days after the adopted  explosion epoch \EpEpoch\\
\end{flushleft}
\end{table}

\end{document}